\documentclass{aa}

\usepackage{graphicx}
\usepackage{txfonts}
\usepackage[]{hyperref} % To add links in your PDF file
\usepackage{booktabs} % for much better looking tables
\usepackage{array} % for better arrays (eg matrices) in maths
\usepackage{paralist} % very flexible & customisable lists (eg. enumerate/itemize, etc.)
\usepackage{verbatim} % adds environment for commenting out blocks of text & for better verbatim
\usepackage{ulem}

\makeatletter
\renewcommand*\aa@pageof{, page \thepage{} of \pageref*{LastPage}}
\makeatother

\newcommand{\crires}{\ensuremath{\mathrm{CRIRES+}}}
\newcommand{\pyreduce}{PyReduce\ }
\usepackage{natbib}
\bibpunct{(}{)}{;}{a}{}{,} % to follow the A&A style
\graphicspath{{./img/}}

\begin{document}

   \title{Optimal extraction of echelle spectra: getting the most from observations}

% Because the world isn't flat
   \subtitle{}

   \author{Nikolai Piskunov\inst{1,2} \and Ansgar Wehrhahn\inst{1} \and Thomas Marquart\inst{1}}

   \institute{Department of Physics and Astronomy, Uppsala University, Box 516, S-75120 Uppsala, Sweden
   \and
   INASAN, 119017, Pyatnitskaya str., 48, Moscow, Russia
   }

   \date{Received \today; accepted \textit{already}?}

% \abstract{}{}{}{}{} 
% 5 {} token are mandatory
 
  \abstract
   {The price of instruments and observing time on modern telescopes is quickly 
   increasing with the size of the primary mirror. Therefore, it is worth revisiting the
   data reduction algorithms to extract every bit of scientific information from observations.
   Echelle spectrographs are typical instruments in high-resolution spectroscopy, but attempts
   to improve the wavelength coverage and
   versatility of these instruments results in a complicated and variable footprint of
   the entrance slit projection onto the science detector. Traditional spectral extraction methods
   fail to perform a truly optimal extraction, when the slit image is not aligned with the
   detector columns but instead is tilted or even curved.}
   {We here present the mathematical algorithms and examples of their application to
   the optimal extraction and the following reduction steps for echelle spectrometers
   equipped with an entrance slit, that is imaged with various
   distortions, such as variable tilt and curvature. The new method minimizes the loss
   of spectral resolution, maximizes the signal-to-noise ratio, and efficiently identifies
   local outliers. In addition to the new optimal extraction we present order splicing
   and a more robust continuum normalization algorithms.}
   {We have developed and implemented new algorithms that create a
   continuum-normalized spectrum. In the process we account for
   the (variable) tilt/curvature of the slit image on the detector and achieve optimal
   extraction without prior assumptions about the slit illumination. Thus the new method
   can handle arbitrary image slicers, slit scanning, and other observational techniques
   aimed at increasing the throughput or dynamic range.}
   {We compare our methods with other techniques for different instruments to illustrate
   superior performance of the new algorithms compared to commonly used procedures.}
   {Advanced modelling of the focal plane requires significant computational effort but
   it pays off by retrieving more science information from every observations. The described
   algorithms and tools are freely available as part of our \pyreduce package.}


   \maketitle
%
%________________________________________________________________
         
\section{Introduction, Motivation, and History}
Nearly 20 years ago one of the authors, together with Jeff Valenti, created an algorithm for the optimal
extraction of echelle spectra without prior assumptions about the cross-dispersion profile. Initially the
math was presented only for the case when the slit image is perfectly aligned with the detector columns, but
a brave promise to extend this in the future to tilted and curved slit images was made. Only 17 years
after the original paper \citep{2002A&A...385.1095P} we are about to deliver on this promise.

The purpose of this paper is to describe the current status of the data processing package REDUCE that became
fairly popular for extracting 1D wavelength-calibrated spectra from the data taken with cross-dispersed echelle
spectrometers. Such instruments combine high efficiency with high spectral resolution, but
the need for angular separation of spectral orders (cross-dispersion) makes data extraction notoriously
difficult, since spectral orders in the focal plane have variable spacing and shape. In slit instruments
the direction of the main dispersion is not perpendicular to the spatial direction (slit image) and in
Non-Littrow optical schemes \citep{Littrow,2009AN....330..574K} the slit images often exhibit variable
tilt and even curvature across the focal plane.
Here we define the "optimality" of the extraction in terms of maximizing the signal-to-noise ratio (S/N)
per resolution element, while preserving the spectral resolution delivered by the optical system. Our
original optimal extraction was presented together with several other algorithms (that now make up
the core of the REDUCE package) in \cite{2002A&A...385.1095P}, that henceforth we will refer to as
PAPER I. We call the optimal extraction algorithm "slit decomposition" as it decomposes a 2D image
registered by the focal plane detector into two vectors: a spectrum and a slit illumination function.
We make no assumptions about the shape of the slit illumination function (also known as the
cross-dispersion profile, see below). The original algorithm had a number of limitations
that took some effort to sort out. The major restriction was the assumption that the spectrometer creates a
rectangular slit image for each wavelength and that it is strictly parallel to the pixel columns on the detector.
Another restriction was the use of IDL as programming language. Yet another was the speed optimization
strategy that was clearly insufficient. At the time of writing of PAPER I certain tools (e.g. the wavelength
calibration and continuum normalization) did not even exist.

%AW: Any other methods that need to be mentioned? What about other implementations of the optimal extraction algorithm, maybe something like this:
Several other implementations of the optimal extraction algorithm exist (e.g. \citealt{1989PASP..101.1032M,2004PASP..116..362C,2008AcASn..49..327C,2014A&A...561A..59Z,2020AJ....159..187P}), but this is the first one to allow for a slit function that is not aligned with the detector columns.
%(e.g., Marsh 1989; Kinney et al. 1991; Hallet al. 1994; Bacon et al. 2001; Cushing et al. 2004; Bershady et al. 2005; Zanichelli et al. 2005; Dixon et al. 2007; Cui et al. 2008, Zechmeister et al. 2014, Petersburg et al 2020)

%AW: This is the new paragraph about other methods, might have to move it further down? I feel like this should be somewhere else either way. But I'm not sure where. Maybe even the discussion at the end? Or in the introduc
%AW: The perfect extraction is also much more computationally expensive

Since the release of PAPER I the "perfect" extraction (also known as "spectro-perfectionism") method has been developed \citep{2010PASP..122..248B,2019PASP..131l4503C}. While the method was designed for the recovery of faint objects, it can also be used in stellar spectroscopy. It is however much more computationally expensive, and suffers from several other practical difficulties. Most importantly the algorithm relies on knowledge of the PSF shape at any given wavelength, which is difficult to determine accurately. This also means that any instrument shift will need to be corrected. %a condition that is difficult to maintain at high resolution (R > 30~000). 
%Another very important aspect of it is that it works best in the absence of continuum, that is the registered spectrum consists of separate and unresolved emission lines sitting on top of some 2D background that is not directly related to the emission of the target. This is not a typical use case for REDUCE and is not considered in this paper.
We note that, REDUCE has no concept of wavelength or PSF created by diffraction (and thus its limitation), but by working with pixels it can easily accommodate drifts between calibrations and science exposures, in both spatial and dispersion directions.

Writing this paper gives us an opportunity to catch up with the development of REDUCE, i.e. to write up the math
behind the main algorithms and speed optimization concepts. It feels good to have it stored for posterity in
one place!

In the next three sections of the paper we will present the algorithms of optimal extraction in the case of a tilted
and even curved slit image (\autoref{sec:method}), the 1D and 2D wavelength calibration methods and their
inter-comparison (\autoref{sec:wavecal}), and the continuum normalization (\autoref{sec:continuum}).

After that we will present the Python and C implementation of the main algorithms as well as some examples
illustrating the performance of the latest REDUCE version in terms quality of the data reduction
(\autoref{sec:implementation}).

\section{Generalized slit decomposition algorithm}
\label{sec:method}
We start this section by reminding the reader, and ourselves, of the algebra required to decompose (a fragment of) a spectral order image created by an echelle spectrograph on a matrix detector. We will represent the image as an external product of two vectors: the s{L}it illumination function $L$ and the s{P}ectrum $P$. In the following our convention is that the main dispersion is \textit{approximately} aligned with the detector rows, while the cross-dispersion or spatial direction \textit{approximately} follows the columns.

In addition to a 2D image of a spectral order sampled on the detector, we also need the trace of the order, i.e. the location of the spectral order in the pixel coordinate system. A robust algorithm for order tracing was presented in PAPER I. The image itself is supposed to be corrected for bias, dark current, and other types of background (e.g. by combining nodding images for IR observations) and flat-fielded. To avoid noise amplification we use the so-called "normalized" flat field (PAPER I). The normalized flat field is the flat field data (usually a master flat) divided by the flat field model
constructed by the slit decomposition algorithm from the cross-dispersion profile and the blaze function. In places where the flat
field signal is low (e.g. between spectral orders) the ratio is set to one to avoid introducing additional noise. Thus the normalized flat contains only positive values around one. It carries the information about relative pixel
sensitivity and is not affected by the signal variation in the original master flat. This is particularly important
for the fiber-fed instruments where flat field signal decreases quickly in cross-dispersion direction.

\begin{figure}
    \centering
    \includegraphics[width=1.\columnwidth]{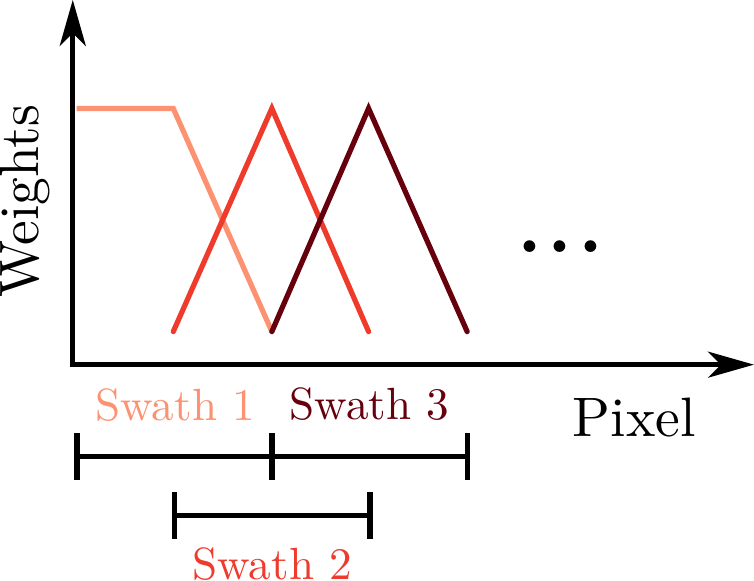}
    \caption{The distribution of the first three swaths in an order. Each swath is overlapped by two other swaths (except for the first and last swath), so that each pixel is contained in two swaths. The swaths are combined into the final spectrum using linear weights with the central pixels having the most weight.}
    \label{fig:swath}
\end{figure}

Usually, we process the whole spectral order in a sequence of overlapping rectangular segments that we call \emph{swaths} (see \autoref{fig:swath} for illustration). Using swaths instead of the whole order allows us to account for changes in the slit illumination function which is held constant within each swath. The shorter the swaths, the more variability in the slit illumination function along the spectral order can be reproduced, replicating optical aberrations and optical imperfections. On the other hand, for low S/N data wider swaths make the decomposition more robust and help extracting the available signal.

\begin{figure}
    \centering
    \includegraphics[width=1.\columnwidth]{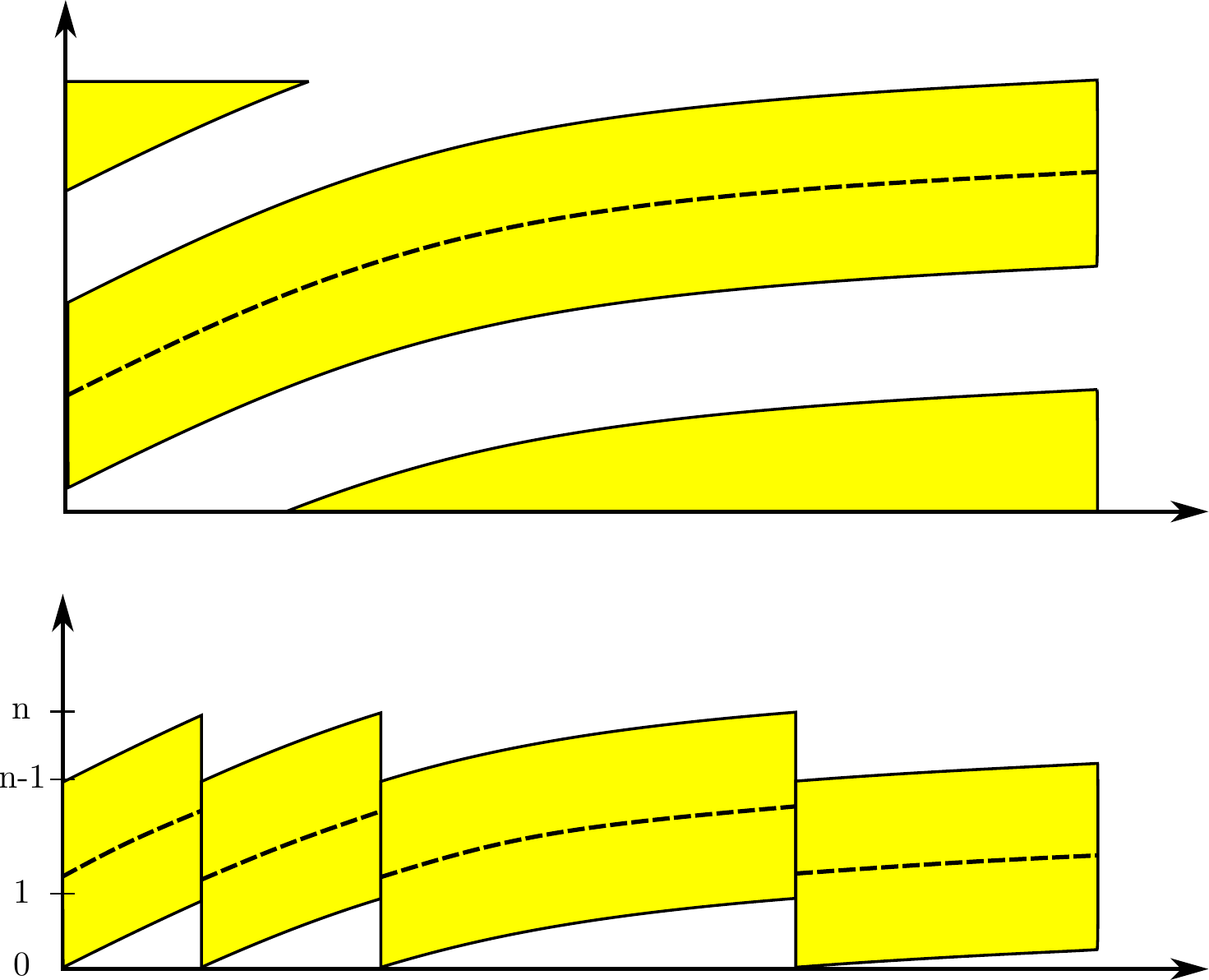}
    \caption{Top: Input Image with selected order (dashed line) and two adjacent orders. Bottom: Image packed for slit decomposition with the discontinuous order trace (dashed line).}
    \label{fig:rectify}
\end{figure}

The vertical (cross-dispersion) extension of the swaths is determined relative to the order trace using two parameters: the number of pixels below and above the trace. This provides practical flexibility in case the image used for order localization has an offset relative to the science image. REDUCE offers a special tool that can estimate the extraction height using the signal level drop in spatial direction in the central swath or by fitting a cross-dispersion profile with a Gaussian. If neither of the two methods is acceptable, the user can also specify the offset in pixels explicitly. The extraction height should cover only the current spectral order. Extending it further will not change the extraction result. This is because the derived signal is actually being computed using a model, which constructed by the slit decomposition and not from the measured pixel counts. One should of course not include adjacent orders and be aware of the increase in computation time. Using a too narrow height may (and will) affect the quality of the model and thus the resulting spectrum as we loose some information. The extraction height selection is illustrated in the top sketch of \autoref{fig:rectify}. Once the central line crosses the pixel row, the initial and final extraction row numbers are shifted accordingly. The vertical size of the swath affects the computational time, so we compress the swath by packing pixels that fall into the extraction range for each column into a rectangular array. Every time the center line crosses the pixel row the new central line and the packed array exhibit discontinuity as shown in the bottom panel of \autoref{fig:rectify}. The order trace line $y_c(x)$ is now contained within a single row of detector pixels. We then choose to shift the order trace to the bottom of the pixel row so that it only has values between 0 and 1.
 
\subsection{Problem Setup}

For an ideal cross-dispersed echelle spectrometer measuring a cosmic (faraway) source the image of a spectral order consists of many monochromatic images of the entrance slit characterized in first approximation by the relative intensity distribution in spatial direction (the slit illumination function). Each slit illumination function is scaled by the relative number of photons in the corresponding wavelength bin (spectrum). Thus, it should be possible to represent a 2D image of a spectral order by two 1D functions and even reconstruct these functions directly from the 2D image registered by the detector pixels.

The goal of the slit decomposition algorithm is to model the 2D intensity distribution in a rectangular area of the detector that contains an image of a spectral order or its fragment. We assume that the detector consists of square, equally-spaced and equally-sized pixels and contains no gaps between pixels. This is not strictly true in practice, but neither is it a limiting assumption: it is easy to introduce the physical coordinates and the size of each pixel and use these instead of pixel numbers and pixel contribution in the point spread function (PSF) footprint. We will not do this here because the following algebra is complex enough even without this extra layer of transformation. Thus in the following we will stick to rows and columns as coordinates. As mentioned before, we assume the main dispersion direction is roughly horizontal, so that the wavelength inside the order changes in $x$-direction, while the spatial extension of the slit is approximately vertical. REDUCE provides the transformation mechanism for achieving this orientation for any given instrument. The input for the problem is the 2D photon count surface measured by the detector $S_{x,y}$ and the trace of the order location $y_c(x)$.

\subsection{Decomposition in case of the strictly vertical 1D PSF}

\begin{figure}[ht]
    \begin{center}
        \includegraphics[width=0.95\columnwidth]{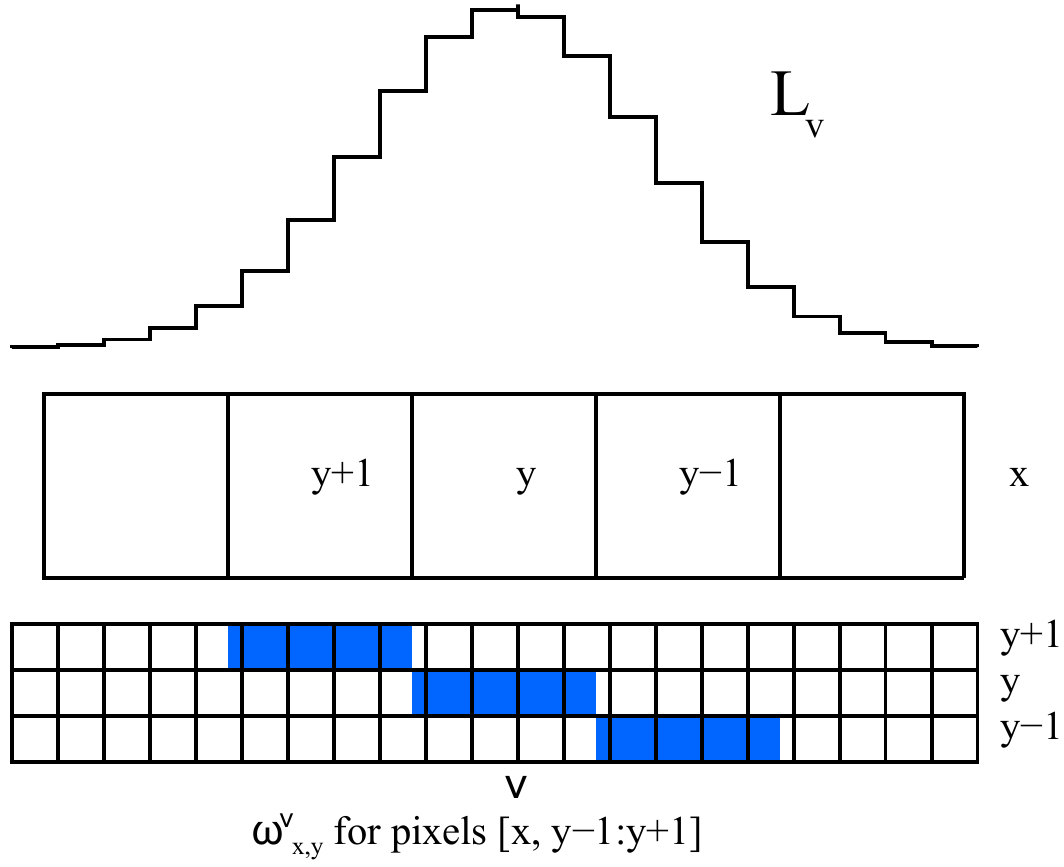}
        \caption{Schematic presentation of the subgrid sampling of the slit illumination function $L$ and the corresponding structure of the $\omega$ tensor. On the top is an example of a slit illumination function aligned for column $x$. $L$ is set on a subpixel scale $v$ (shown at the bottom panel) that may shift relative to the detector pixels from one column to the next. For three pixels in this column centred at row $y$ we show the structure of the corresponding section of $\omega_{x,y}^v$. The fraction of each subpixel filled with blue is proportional to the value of the corresponding element of $\omega$.}
        \label{subpixel grid 1D}
    \end{center}
\end{figure}

We start by assuming an ideal case where the monochromatic images of the slit are perfectly aligned with the columns of the detector. Then our model for the photon count $S_{x,y}$ on detector pixel $(x,y)$ is given by an outer product of the two vectors: a continuous slit-illumination function $L(v)$ and the discrete spectrum $P_x$:

 \begin{equation}
     \label{eq model}
     S_{x,y} = P_x\cdot \int_{y-y_c(x)}^{y-y_c(x)+1} L(v) dv
 \end{equation}
 
 where we assume that the central line of the order $y_c(x)$ is known precisely. $L$ drifts vertically across pixels from one column to the next due to the tilt of the spectral order. This is described by the shift of integration limits relative to the detector row $y$. The value of the shift is given by the order trace $y_c(x)$. The slit illumination function $L$ remains independent of $x$ on the $v$ grid. The spectrum $P$ changes from one column to the next. To avoid scaling degeneracy between $L$ and $P$ we postulate that the area under $L$ should be equal to 1. For IR instruments special care should be taken when using chopping/nodding techniques to avoid the effect of the negative values. Normally, electronic detectors do not generate negative signal. The background signal is removed by subtracting the bias correction.%This is the purpose of the bias correction. %AW: the formulation was awkward so I removed the mention of the bias correction
 When subtracting images within chopping/nodding pairs however one can still get negative values. Preserving these values is important to not distort the noise distribution function (see e.g. \cite{2005A&A...443.1087L}).

 In practice Equation~\ref{eq model} will not hold precisely even if all our assumptions are met. This is due to noise in the observation, ghosts and scattered light in the spectrometer, cosmic rays, detector defects etc. Thus, our model will fit the measurements only approximately and in some pixels it will not fit at all. This means that the model cannot be constructed for each pixel individually, but has to be derived from a segment of the image in some sort of least squares sense. To do that the slit function must be set on some discrete grid. This grid must be finer than the pixel size if we want to account for a smooth shift of the central line and for a potentially complex structure of $L$. We create such a grid by introducing an integer oversampling factor $O$, so that $1/O$ gives the step of the fine grid (or subgrid) in units of detector pixel size. Now we can express the requirement for the model to match the data as a least squares minimization problem:
 \begin{equation}
     \label{eq fit}
    \Phi=\sum_{x,y} \left[E_{x,y} -  S_{xy}\right]^2=
    \sum_{x,y} \left[E_{x,y} -  P_x\cdot\sum_v \omega_{xy}^v\cdot L_v\right]^2 = \mathrm{min}
 \end{equation}
 where $E_{xy}$ are the actual measurements. Tensor $\omega_{x,y}^v$ is proportional to the fraction of the area of subpixel $[x,v]$ that falls inside the detector pixel $[x,y]$. That is, for a detector pixel $[x, y]$, $\omega_{x,y}^v$ is equal to $1/O$ for all $v$ that are fully contained inside this detector pixel, less than $1/O$ for the two boundary values of $v$ and 0 for all other indices. Note, that due to our selection of the subgrid the sum of the two boundary values for every $x$ and $y$ is also $1/O$. \autoref{subpixel grid 1D} illustrates the properties of $\omega_{x,y}^v$.

The tensor $\omega$ is the key for representing the projection of the monochromatic slit images onto detector pixels with only a slight generalisation needed to follow a tilted or even curved slit, which we will describe in the sections below. Here we deal with a strictly vertical slit projection and thus can note a few important properties of $\omega$ that will be the basis for speed optimisations here and later on. Each detector pixel is sampled by a maximum of $O+1$ subpixels of $\omega$ that may have non-zero values. Out of these, all intermediate elements are equal to $1/O$. The first and the last elements are less or equal to $1/O$, but their sum is equal to $1/O$, so that the integrated weight for any detector pixel $x,y$ given by $\sum_v\omega_{xy}^v$ is equal to 1. There is also a relation between elements of $\omega$ for two consecutive values of $y$: they overlap by one element in $v$ and the sum of these two elements is again equal to $1/O$.

Now we are going to solve Equation~\ref{eq fit}. First we take the two partial derivatives of this equation over the elements of $P_x$ and $L_v$:
\begin{eqnarray}
    \frac{\partial\Phi}{\partial L_v}&=&-2\sum_{x,y} \left[E_{x,y} -  P_x\cdot\sum_v \omega_{xy}^v\cdot L_v\right]
    P_x\omega_{xy}^v =0\\
    \frac{\partial\Phi}{\partial P_x}&=&-2\sum_{x,y} \left[E_{x,y} -  P_x\cdot\sum_v \omega_{xy}^v\cdot L_v\right]
    \sum_v\omega_{xy}^v L_v=0
\end{eqnarray}

These can be re-written as linear equations for $L_v$ and $P_x$:
\begin{eqnarray}
    \sum_{v'}\left[\sum_{xy}P_x^2\omega_{xy}^v\omega_{xy}^{v'}\right]L_{v'}&=&\sum_{xy}E_{xy}P_x\omega_{xy}^v
    \label{Lv full equation}\\
    P_x\sum_y\left[\sum_v\omega_{xy}^vL_v\right]^2&=&\sum_y E_{xy}\sum_v\omega_{xy}^vL_v
    \label{Px full equation}
\end{eqnarray}
or:
\begin{eqnarray}
    \sum_{v'}{A_{vv'}}L_{v'} = {b_v} \label{Lv equation} \\
    P_x=\frac{\sum_y E_{xy}\sum_v\omega_{xy}^vL_v}{\sum_y\left[\sum_v\omega_{xy}^vL_v\right]^2}
    \label{Px equation}
\end{eqnarray}
where the expressions for the matrix ${A_{vv'}}$ and the right-hand-side (RHS) ${b_v}$ are given by Equation~\ref{Lv full equation}.

Equations~\ref{Lv equation} and \ref{Px equation} are linear but they cannot be combined to separate the unknowns (see also \citealt{1986PASP...98..609H}). Therefore we adopt an iterative scheme alternating between solving 
\autoref{Lv equation} and \autoref{Px equation}. Level of changes in $P_x$ can be used as convergence test.

Note, that the equation for the the spectrum $P_x$ includes the slit function $L_\nu$ as a weight.
This property is responsible to maximizing the S/N in our algorithm.

The whole procedure can be integrated with a "bad pixel" mask that can be dynamically adjusted during the iterations. Suppose $M_{xy}$ is 1 for "good" detector pixels and 0 otherwise. We can rewrite the expressions for matrix ${A}$ and the RHS in Equation~\ref{Lv equation} as:
\begin{eqnarray}
    {A_{vv'}}&=&\sum_{xy}M_{xy}P_x^2\omega_{xy}^v\omega_{xy}^{v'}\label{eq A matrix} \\
    {b_v}&=&\sum_{xy}M_{xy}E_{xy}P_x\omega_{xy}^v\label{eq RHS}
\end{eqnarray}

The equation of $P_x$ will be:
\begin{equation}
    P_x=\frac{\sum_y M_{xy} E_{xy}\sum_v\omega_{xy}^vL_v}{\sum_y M_{xy}\left[\sum_v\omega_{xy}^vL_v\right]^2}
    \label{eq P expression}
\end{equation}
Massive defects, such as bad columns, must be detected beforehand to avoid divisions by zero. In the iteration loop above one can implement adjustments of $M_{xy}$ by constructing the standard deviation between the data and the model as given by Equation~\ref{eq fit} using the current bad pixel map and then correcting the map by comparing the actual difference with the standard deviation.

\subsection{Convergence, selection of oversampling and regularization}

The iterative scheme presented above has excellent convergence properties: typically the unknown functions are
recovered to a relative precision of 10$^{-5}$ in 3 to 5 iterations. The convergence rate besides general
consistency between the data and model depends on the selection of $O$, which deserves a separate discussion.
The oversampling is required to adequately describe the gradual shift of the central line relative to the pixel
rows of the detector and possible features of the slit illumination. Qualitatively, one would expect $O=1$ should be sufficient when a spectral order is strictly parallel to pixel rows. On the other hand, if the central line shifts
by 0.5 pixel over the whole swath, then $O$ could perhaps be 2. The problem is that no cross-disperser (the
low-dispersion spectrometer used to separate echelle orders) keeps spectral orders in straight lines.
This makes it impossible to use a single oversampling value for the whole order. The issue can be alleviated by
selecting $O$ to match the largest tilt while regularizing $L_v$. One suitable form of regularization is a
constraint on the first derivatives (classical Tikhonov regularization, \citealt{Tikhonov1977})
that would damp oscillations of the oversampled slit function. The use of regularization decouples
the selection of the oversampling
factor from the exact order geometry. Similarly, one may want to have an option to control the smoothness of the
spectrum sacrificing its spectral resolution. Such an option is helpful when decomposing the flat field or other
sources where no sharp spectral features are expected. Both regularizations can be easily incorporated into
Equation~\ref{eq fit}:
\begin{eqnarray}
    \label{eq reg fit}
    \Phi&=&\sum_{x,y} M_{xy}\left[E_{x,y} -  P_x\cdot\sum_v \omega_{xy}^v\cdot L_v\right]^2 + \\
    &+&\Lambda_L\sum_v \left(L_{v+1}-L_v\right)^2+\Lambda_P\sum_x \left(P_{x+1}-P_x\right)^2\nonumber
 \end{eqnarray}
 where $\Lambda_L$ and $\Lambda_P$ are the regularization parameters for the two unknown vectors. The corresponding changes to the matrix ${A_{vv'}}$ will affect the main diagonal: $2\Lambda_L$ will be added to all elements except the first and the last that only get one additional $\Lambda_L$. Also $\Lambda_L$ should be subtracted from all elements on the upper and lower subdiagonals. Equation~\ref{Px full equation} will become a tri-diagonal system of linear equations with $\sum_yE_{xy}\sum_v\omega_{xy}^vL_v+2\Lambda_P$ for all $x$ except the first and the last elements where the expression is $\sum_yE_{xy}\sum_v\omega_{xy}^vL_v+\Lambda_P$. All subdiagonal elements will contain $-\Lambda_P$. Note, that the use of regularization for the spectrum is purely optional while
 setting $\Lambda_L$ to zero will most probably lead to a zero determinant of matrix $A_{vv'}$ for $O>>1$.
 
The choice of regularization parameters $\Lambda_L$ and $\Lambda_P$ depends on the S/N of the
 data, the oversampling parameter $O$, as well as the shape of the slit illumination function. For
 reasonable S/N (above 20) it is sensible to set $\Lambda_P=0$ and select $\Lambda_L$ as the
 smallest number that still damps non-physical oscillations in the slit function. Fortunately the extracted spectrum is not very sensitive to the choice of $\Lambda_L$. When
 investigating this issue using ESO UVES, HARPS and CRIRES+ instrument data with S/N$\approx$50
 we discovered that spectra extracted with the best $\Lambda_L$ and $10\times\Lambda_L$ differ by
 less than 0.05\%.
 
\begin{figure}[ht]
    \begin{center}
        \includegraphics[width=0.98\columnwidth]{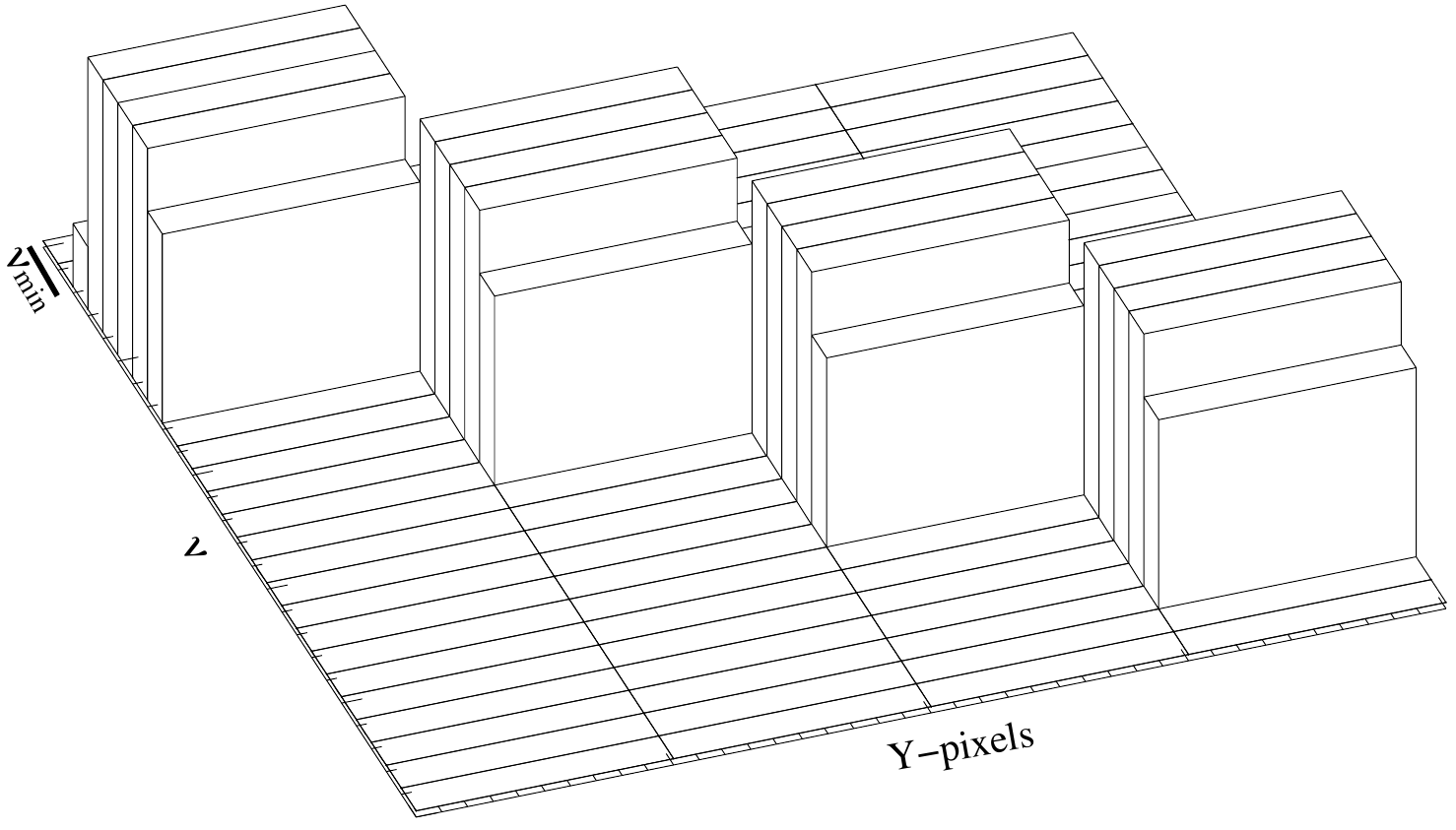}
        \caption{A surface view of a 2D $\omega_{x,y}^v$ tensor projection for a fixed $x$. Horizontal
        axis corresponds to the detector pixel $y$-coordinate. The other coordinate is the
        oversampling direction ($\nu$).}
        \label{omega lego}
    \end{center}
\end{figure}

\begin{figure}[bt]
    \begin{center}
        \includegraphics[width=0.95\columnwidth,]{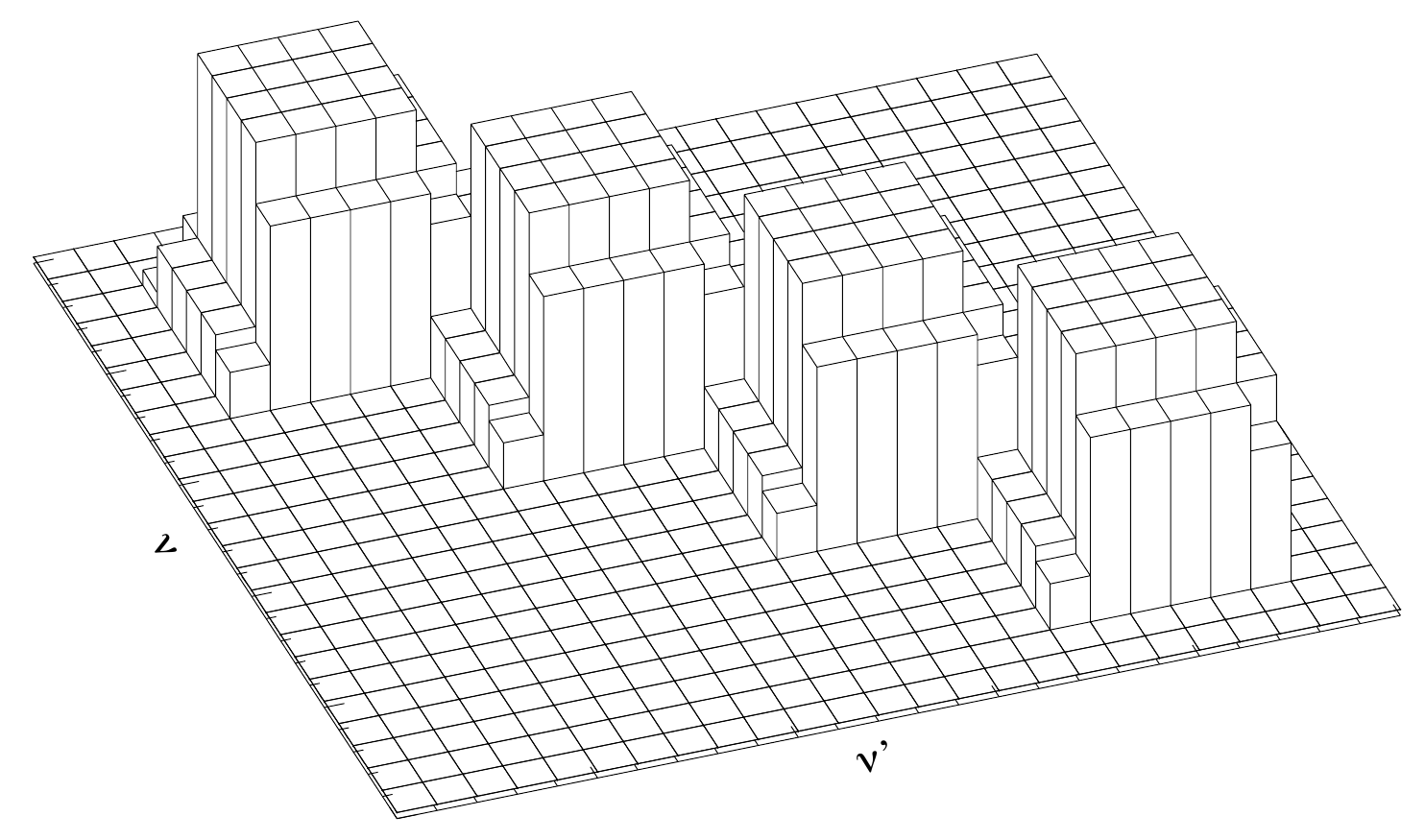}
        \caption{A surface view of $\omega^\top\times\omega$ projection for a fixed $x$. The result is a
        symmetric square matrix with dimensionality of $\nu$.}
        \label{oo lego}
    \end{center}
\end{figure}

\begin{figure}[ht]
    \begin{center}
        \includegraphics[width=0.95\columnwidth, trim=0cm 4.5cm 0 4.cm]{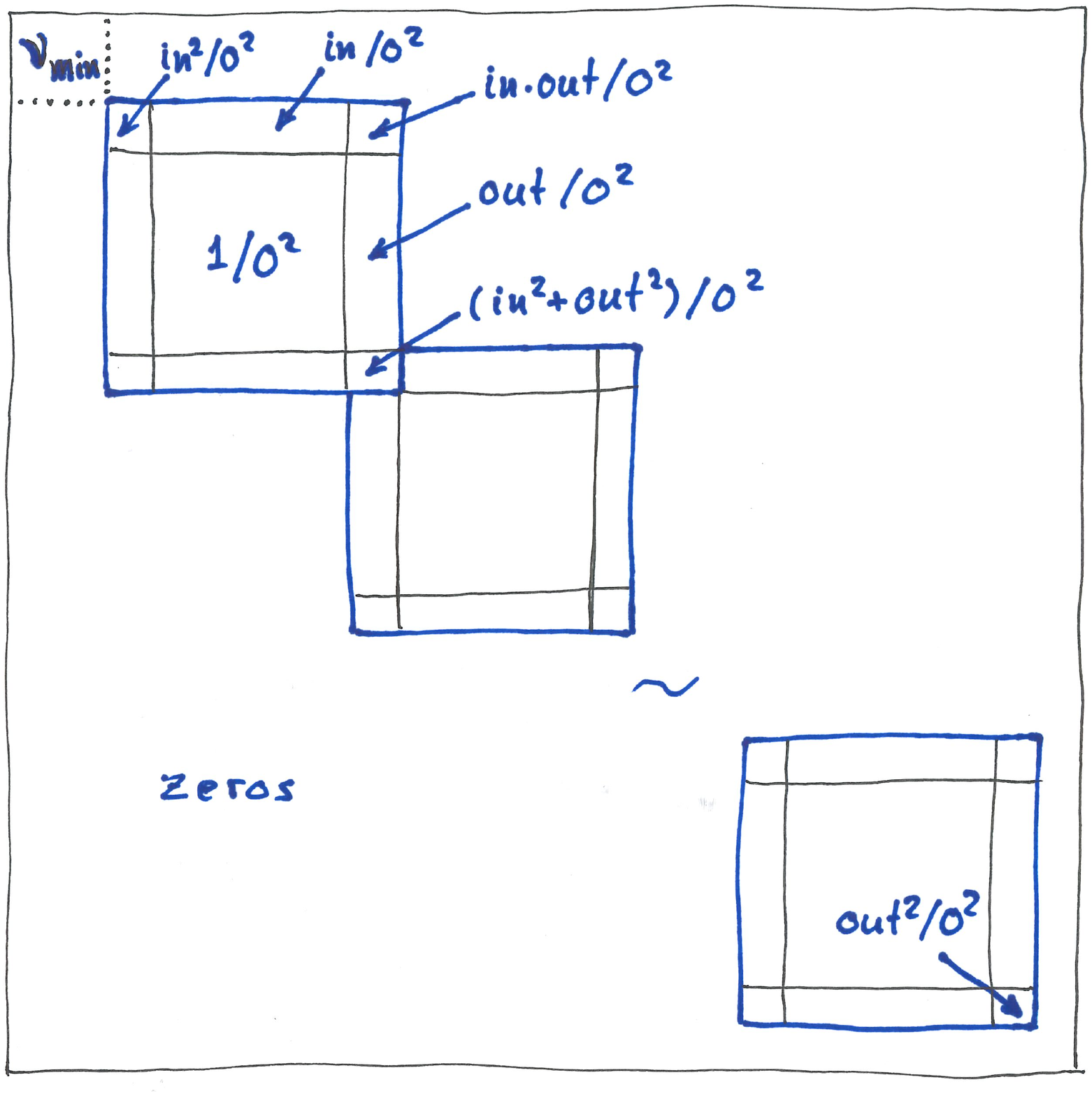}
        \caption{Schematic view of the product of $\omega^T\omega$ for a fixed value of $x$. The
        result is a squared and symmetric matrix of order $(N_y+1)\times O+1$ where $N_y$
        is the height of the packed swath. The outer side of the "squares" on the main diagonal
        is $O+1$. The squares overlap by one row/column. The offset from the top left corner is
        determined by the central line: a $y_c$ of zero will imply $v_{min}=0$. The $in$ and
        $out$ elements are the footprints of the first and the last subpixels that fall in a given
        detector pixel $x,y$. A central line offset of zero would set $in=1$ and $out=0$.
        All elements outside the main diagonal boxes are zero. The values inside a box are known
        explicitly as shown on the sketch.}
        \label{oo sketch}
    \end{center}
\end{figure}

\subsection{Optimisation in case of vertical slit decomposition}

Now that the actual slit decomposition is reduced to repeatedly solving a system of two linear equations, we can examine the performance. The typical size of the final systems are given by the packed height of a swath times the oversampling $O$ (typical numbers are 30$\times$10) for $L_v$, and the width of a swath (typically between 200 and 800 columns) for $P_x$. The main complication is the construction of the matrices involved. This process involves the multiplication of a substantially larger tensor $\omega_{xy}^v$ with itself and with $P_x$. Note, that $\omega_{xy}^v$ describes the geometry of the spectral order and thus remains constant throughout the iterations for a given swath. That offers two paths for efficient construction of the matrices involved in Equations~\ref{eq A matrix}-\ref{eq P expression}.

The first path is to reduce the size of the largest summation using the structure of the $\omega$ tensor. Constructing ${A_{vv'}}$ (\autoref{eq A matrix}) is by far the most expensive part of an iteration, but a major part can be pre-computed knowing the order trace line. This part is $\sum_y \omega_{xy}^v \omega_{xy}^{v'}$. For a given column $x$ the 2D projection of $\omega_{xy}^v$ has a layout similar to the example presented in Figure~\ref{subpixel grid 1D}. Note the self-similar pattern that shifts by $O$ subpixels when moving to the next pixel. A product of two such matrices (that is $\sum_y \omega_{xy}^v \omega_{xy}^{v'}$ for a fixed $x$) on a $vv'$ plane can be evaluated analytically as explained below. Figure~\ref{omega lego} shows a typical layout of the $\omega$ projection for a fixed $x$. The structure is self-similar: for each $y$-column the first non-zero element (from the top of the image) corresponds to $L$ entering the $x,y$ pixel, followed by a set of $O-1$ elements in $\nu$ with identical $1/O$ values. The sequence finishes with the last value corresponding to leaving pixel $y$. For the first $y$ (on the left) the pattern is offset by $\nu_\mathrm{min}$ from the top by the central line $y_c(x)$. The next $y$ will have the same pattern offset by an additional $O$ in $\nu$, etc.

We name $in$ and $out$ the fractions of the first and the last subpixel of the slit function (referenced with $\nu$) that fall into detector pixel $y$ in column $x$. $in$ and $out$ have values between 0 and 1 and their sum is precisely $1$, because we choose $O$
to be an integer number.

The matrix product of $\omega^\top \times \omega=\sum_y \omega_{xy}^v \omega_{xy}^{v'}$ is graphically
presented in Figure~\ref{oo lego}. This symmetric matrix has repeating square structures around the
main diagonal starting at $\nu_\mathrm{min}+1$. The side of each square has a length of $O+1$.
Surprisingly, this matrix contains only 7 unique, non-zero elements: two in the corners of the
square blocks located on the main diagonal
($in\cdot out/O^2$ and $(in^2 + out^2)/O^2$), two in the middle part of the horizontal and the vertical
border of each square ($in/O^2$ and $out/O^2$), one in the middle of each square ($1/O^2$), and two in
the first and the last non-zero diagonal elements, that are different from the square overlap pixels
($in^2/O^2$ and $out^2/O^2$) (see Figure~\ref{oo sketch}). The squares on the main
diagonal overlap by one element. The value of all these pixels is the same: $(in^2 + out^2)/O^2$. The upper
left corner of the first square and the bottom right corner of the last square do not overlap with
anything, so their values are $in^2/O^2$ and $out^2/O^2$ correspondingly. Equipped with this knowledge
we can optimise the slit decomposition iterations in the following way:

\begin{enumerate}
    \item[1$^\circ$]  Construct the initial guess for the spectrum by e.g. collapsing the input image in the cross-dispersion direction.
    \item[2$^\circ$]  Construct matrix ${A_{vv'}}$ as given by \autoref{eq A matrix}. At this point we will use the insights of this section to generate the product of the two $\omega$'s in the left-hand-side;
    \item[3$^\circ$]  Evaluate the right-hand-side and solve the Equation~\ref{Lv full equation} for $L_\nu$; Normalize the result, setting the integral of $L_\nu$ to 1;
    \item[4$^\circ$]  For each $x$ compute $\omega_{xy}^\nu$ multiplied by the slit function $L_\nu$. Use the product to solve for the spectrum $P_x$ according to Equation~\ref{Px equation};
    \item[5$^\circ$]  Evaluate the model image as $P_x\cdot\sum_\nu \omega_{xy}^\nu\cdot L_\nu$ as in \autoref{eq model}. Compare the model with the input image $E_{xy}$, find outliers and adjust the mask;
    \item[6$^\circ$]  Iterate starting from 2$^\circ$ until the change in the spectrum is less than a given margin.
\end{enumerate} 
Note, that the iterations require re-calculations of neither $\omega_{xy}^\nu$ nor of its product $\sum_y \omega_{xy}^v \omega_{xy}^{v'}$.

\subsection{Alternative Optimisation Strategy}
The alternative optimisation approach is based on storing the contributions of every slit function element
to a given detector pixel, and every detector pixel to a given slit function element. The former tensor is
actually very similar to $\omega_{xy}^v$. We will call it $\xi_{x}^{v,n}$. The subscripts have the usual
meaning and the superscript $n$ can take one of the two values: $L$ (Lower) or $U$ (Upper), corresponding to the
cases when an element of the slit function falls onto the boundary of a detector pixel. Each element of tensor
$\xi$ has a composite value (a structure). For every combination of indices it contains the pixel row number
$y$, to which subpixel $v$ contributes (we will write it as $\xi_{x}^{v,n}.y$), and the contribution
value (footprint) that is between
$0$ and $1$, written as $\xi_{x}^{v,n}.w$. One can see that $\xi$ carries the same information
as $\omega$, but it is much more compact as we avoided storing most of the zeros. $\xi$ needs a counterpart
that we will call $\zeta$. Each element $\zeta_{x,y}^m$ carries the information about all elements of the slit
function affected by detector pixel $xy$. The index $m$ runs a range between $1$ to $O+1$ in order
to account for the maximum number of contributing subpixels. Similar to $\xi$, $\zeta$ carries two values:
the number of the slit function elements $v$ referred to as $\zeta_{x,y}^m.v$ and its contribution to this
element $\zeta_{x,y}^m.w$, which is normally $1/O$ except for the boundary subpixels and top/bottom rows
of the swath. Note, that both new tensors are, like $\omega$, only functions of order geometry and thus
need to be computed only once. The purpose of these tensors becomes clear once we rewrite \autoref{eq fit}
and its derivatives with their help:
\begin{eqnarray}
S_{xy}&=&P_x \sum_m L_{\zeta_{x,y}^m.v}\zeta_{x,y}^m.w\\
\Phi &=& \sum_{xy} \left[E_{xy}- P_x \sum_m L_{\zeta_{x,y}^m.v}\zeta_{x,y}^m.w\right]\\
\frac{\partial \Phi}{\partial P_{x}}& = &\sum_{v,n} \left(E_{x,\xi_{x}^{v,n}.y}-S_{x,\xi_{x}^{v,n}.y}\right)\cdot
     L_{v}\xi_{x}^{v,n}.w = 0.\\
\frac{\partial \Phi}{\partial L_{v}} & =& \sum_{x,n} \left(E_{\xi_{x}^{v,n}.x,\xi_{x}^{v,n}.y}-
S_{\xi_{x}^{v,n}.x,\xi_{x}^{v,n}.y}\right)\cdot P_{x}\xi_{x}^{v,n}.w = 0.
\end{eqnarray}
What happens is that the summation is carried out essentially over the non-zero elements of
$\omega^T\omega$ only. The speed-up can be estimated from \autoref{oo sketch} as the ratio
of the number of non-zero elements $N_y\times (O+1)^2-N_y+1$ to the total number $(N_y+1)^2\times O+1)^2$.
In practice, for a packed swath height of 20 we see a bit more than factor 20 in performance increase
compared to the direct construction of matrices involved in linear equations.

The first optimisation path, based on the analytical construction of the $\omega^T\omega$ results in an
even better performance (gain of around another 20\,\% at the expense of larger memory use), but unlike
the 2\textsuperscript{nd} path its advantage vanishes, when we introduce a bad pixel mask 
as in Equations~\ref{eq A matrix}, \ref{eq RHS}, and \ref{eq P expression}. The mask is involved in computing
the product of $\omega^T\omega$, forcing the re-computation of this product during every iteration.
Thus we will not use this approach for the case of a tilted or curved slit image.

\subsection{Decomposition in case of a curved/tilted 1D PSF}
\label{ssec:tilt_decomp}

\begin{figure}[ht]
\begin{center}
\includegraphics[width=0.99\columnwidth]{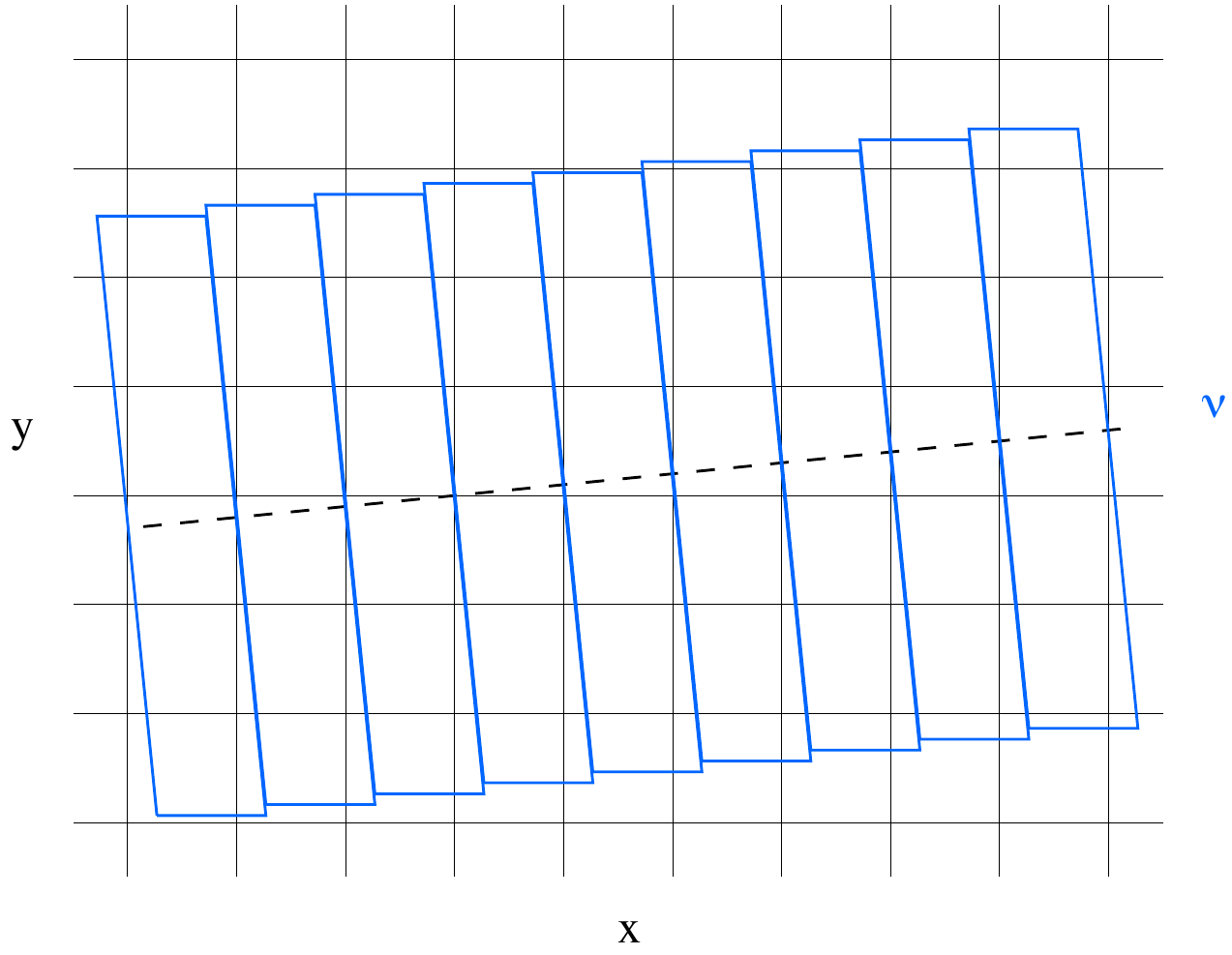}
\caption{Schematic view of "ideal" monochromatic images of a slit projected onto a detector by a
non-Littrow spectrometer. Black squares represent spectrometer pixels. The dashed line
traces the center of a spectral order and the blue boxes show the idealised footprints of slit images.
Note, that by design the center of the spectral order and the sides of a slit image intersect at the
pixel column boundaries of the detector.}
\label{tilted slit}
\end{center}
\end{figure}

In this section we explore the decomposition of a 1D slit bent by a known amount relative to the
detector columns. Assume thus that the offset of monochromatic images of the slit from a vertical
line on the detector is described by a second order polynomial:
\begin{equation*}
  \delta x(\delta y) = a(x)\cdot \delta y^2 + b(x)\cdot \delta y +c(x),
\end{equation*}
where $\delta y = \mathbf{y - y_c(x)}$ with $y_c(x)$ being again the central line of the order on the detector.
We postulate that the offset from column $x$ at the level of the order trace $y=y_c(x)$ is zero.
This means that $c(x)$ is always 0. In the case of a strictly vertical
slit image the offset expression is reduced to a trivial $\delta x(\delta y) = 0$. For a straight
but tilted slit image we will have only the linear term: $\delta x(\delta y) =b(x)\cdot \delta y$.
The presence of a horizontal offset means that subpixels may now contribute to adjacent columns, changing the
structure of the $\omega$ tensor. Therefore the model for detector pixel $x,y$ should be modified to reflect this:
 \begin{equation}
 \label{eq curved slit model}
S_{x,y} = \sum_{ix=-\Delta x}^{+\Delta x}P_{x+ix}\cdot\sum_{iy} \omega_{x+ix,y}^{-ix,iy}\cdot L_{iy},
 \end{equation}
where $\Delta x = \lceil \delta x \rceil$ is the pixel range affected by the curvature.
Compared to the previous section we are now using the index name $iy$ to address the elements of $L$ instead
of $v$ to emphasize the difference between offsets in $x$ and $y$ directions.

Tensor $\omega$ acquired an extra dimension to reflect the contribution to the detector column(s) adjacent to $x$.
In practice, the size of $\omega$ does not have to increase dramatically (typically by a factor of 3) since 
the height of the slit illuminated by a point source is typically small and even noticeable slit curvature
will not result in a large offset in dispersion direction. In long-slit observations one would not want to use slit
decomposition to keep the spatial information. Notable exceptions are the use of an image slicer with many
(5-7) slices or an IR spectrum that is a combination of the two nodding positions. 
Note also, that the column index $x$ of the generalized $\omega$ follows the spectrum $P$, but since we are
interested in the contribution of $P_{x+ix}$ to the pixel $x$ in \autoref{eq curved slit model}, the offset
index $ix$ has the opposite sign to the difference between the contributing column ($x+ix$) and the column
($x$) (the column, for which the model is constructed).

Partial derivatives of the model $S_{xy}$ are:
 \begin{eqnarray}
 \frac{\partial S_{x,y}}{\partial P_{x'}} & = & \sum_{iy} \omega_{x',y}^{x-x',iy}\cdot L_{iy} \\
 \frac{\partial S_{x,y}}{\partial L_{jy}} & = & \sum_{ix=-\Delta x}^{+\Delta x}P_{x+ix}\omega_{x+ix,y}^{-ix,iy}
 \label{eq curved model derivatives}
 \end{eqnarray}

As for the vertical slit case we can formulate the optimisation problem for matching the model
to (a fragment of) a spectral order $E_{x,y}$:
\begin{equation}
\Phi=\sum_{x,y} \left(S_{x,y}-E_{x,y}\right)^2=min,
\end{equation}  
Note however, that there is no more a one-to-one correspondence between the measured swath and the data
needed for the model. This is obvious from \autoref{tilted slit}: if the black squares represent the selected
swath then clearly the model of the first and last columns requires spectrum $P_x$ values that cannot be reliably
derived from this swath (partial slit images). Overlapping zones would solve this problem by carrying
the values of $P_x$ from one swath to the next.

The necessary condition for a (local) minimum is the first derivatives being zero:
 \begin{eqnarray}
 \frac{1}{2}\frac{\partial \Phi}{\partial P_{x'}} & = & \sum_{x,y} \left(S_{x,y}-E_{x,y}\right)\cdot
   \frac{\partial S_{x,y}}{\partial P_{x'}} = 0\\
 \frac{1}{2}\frac{\partial \Phi}{\partial L_{iy}} & = & \sum_{x,y} \left(S_{x,y}-E_{x,y}\right)\cdot
   \frac{\partial S_{x,y}}{\partial L_{iy}} = 0
 \end{eqnarray}
 
Substituting Equations~\ref{eq curved slit model} - \ref{eq curved model derivatives} into the last
two equations we get systems of linear equations for $P$ and $L$:
 \begin{eqnarray}
\label{eq curved slit eq for P and L}
 \sum_{x=x'-\Delta x}^{x'+\Delta x} \sum_y \left(\sum_{ix=-\Delta x}^{+\Delta x}P_{x+ix}\cdot
 \sum_{iy} \omega_{x+ix,y}^{-ix,iy}\cdot L_{iy}-E_{x,y}\right) \cdot  \nonumber \\
 \cdot \sum_{iy} \omega_{x',y}^{x-x',iy}\cdot L_{iy} & = & 0\\
 \sum_{x,y} \left(\sum_{ix=-\Delta x}^{+\Delta x}P_{x+ix}\cdot\sum_{iy} \omega_{x+ix,y}^{-ix,iy}
 \cdot L_{iy}-E_{x,y}\right) \cdot \nonumber\\
 \cdot\sum_{jx=- \Delta x}^{+\Delta x}P_{x+jx}\omega_{x+jx,y}^{-jx,jy} & = & 0
  \end{eqnarray}

Now we are going to re-organise Equation~\ref{eq curved slit eq for P and L} by first substituting $x$
with $x'+jx$, then dropping "prime", and finally shifting the measured data part to the right-hand side:
 \begin{eqnarray}
\sum_{jx=-\Delta x}^{+\Delta x} \sum_y \sum_{ix=-\Delta x}^{+\Delta x} P_{x+ix+jx}\cdot\sum_{iy} \omega_{x+ix+jx,y}^{-ix,iy}\cdot L_{iy}\cdot  \sum_{jy} \omega_{x,y}^{jx,jy}\cdot L_{iy} \nonumber \\
=\sum_{jx=-\Delta x}^{+\Delta x} \sum_y E_{x+jx,y}\cdot  \sum_{jy} \omega_{x,y}^{jx,jy}\cdot L_{iy}
\label{eq curved slit eq for P}
 \end{eqnarray}
 Renaming $x'$ means that the derivative was taken over $P_x$ rather than over $P_{x'}$.
 
 Finally, we note that Equation~\ref{eq curved slit eq for P} is actually a system of $N_{cols}$  linear
 equations numbered by the value of $x$. The matrix for the system is band-diagonal but not symmetric
 with the width of the band equal to $4\cdot\Delta x+1$.

The system of equations for $L$ is derived in a similar way:
 \begin{eqnarray}
\sum_{iy} L_{iy} \sum_{x,y} \sum_{ix=-\Delta x}^{+\Delta x}P_{x+ix}\cdot \omega_{x+ix,y}^{-ix,iy} \cdot
\sum_{jx=- \Delta x}^{+\Delta x}P_{x+jx}\cdot \omega_{x+jx,y}^{-jx,jy} = \nonumber \\
= \sum_{x,y} E_{x,y}\cdot \sum_{jx=- \Delta x}^{+\Delta x}P_{x+jx}\cdot \omega_{x+jx,y}^{-jx,jy}
\label{eq curved slit eq for L}
  \end{eqnarray}

In this case the matrix of the system is fully filled but symmetric.

\subsection{Decomposition optimisation in case of the curved slit}

The optimisation will follow the second path presented for the vertical slit case. 

Again we define two sets of tensors. One (similar to $\omega$) describes the contribution(s) of subpixel
$iy$, associated with the spectrum centered on detector pixel $x$, to other detector pixels. As before we
name it $\xi$. As before it has three indices, but now the second superscript runs through four options,
reflecting the cases when a subpixel is projected onto the intersection of four detector pixels.
% $x$ is connecting $\xi$ to $P_x$ and $iy$ is the subpixel number.
With the slit image no longer aligned with the detector columns, a subpixel can project onto two columns and occasionally
on a boundary between rows. Thus subpixel $iy$ can have a footprint in two or even four detector pixels, which
are referenced as $\xi_{x}^{iy,[LL/LR/UL/UR]}$. $LL$, $LR$, $UL$ and $UR$ refer to the affected detector pixel
location (lower-left, lower-right, upper-left, or upper-right) relative to the selected subpixel.

For each combination of indices the value of $\xi_{x}^{iy,m}$ is a structure containing
$\{x',y',w\}$. $\{x',y'\}$ are the coordinates of the affected detector pixel and $w$ is the footprint of
subpixel $\{x,iy\}$ inside pixel $x',y'$. Tensor $\xi$ will also be very useful when evaluating partial
derivatives.

The other tensor $\zeta$ is in some sense the inverse of $\xi$. It also has three indices, $\zeta_{x,y}^m$,
where $m$ indexes the contributing subpixels and has a range between $0$ and $(O+1)\times 2$. For each
combination of indices the value of $\zeta$ is a structure containing the two coordinates of the
contributing subpixel and its weight (footprint) $\{x',iy',w\}$.

Equipped with these new tensors we can rewrite the expressions for $S_{x,y}$ as well as the derivatives
of $\Phi$.

Our model for detector pixel $x,y$ can now be expressed as:
\begin{equation}
  S_{x,y} = \sum_m P_{\zeta_{x,y}^m.x'}\cdot L_{\zeta_{x,y}^m.iy'}\cdot {\zeta_{x,y}^m.w}
  \label{eq curved optimized model}
\end{equation}
Note, that unlike \autoref{eq curved slit model} we ended up with a single summation.

For partial derivatives over $P_x$ we keep only pixels receiving a contribution from the slit
image centered on the $\{x,y_c(x)\}$ position on the detector.
\begin{equation}
\frac{1}{2}\frac{\partial \Phi}{\partial P_{x}}  =  \sum_{iy,n} \left(S_{\xi_{x}^{iy,n}.x',\xi_{x}^{iy,n}.y'}-E_{\xi_{x}^{iy,n}.x',\xi_{x}^{iy,n}.y'}\right)\cdot L_{iy}
\xi_{x}^{iy,n}.w = 0
  \label{eq curved optimized derivative over p}
\end{equation}

The analogous expression for the derivatives over $L_{iy}$ is also easily written with the help of tensor $\xi$: 
\begin{equation}
\frac{1}{2}\frac{\partial \Phi}{\partial L_{iy}}  =  \sum_{x,n} \left(S_{\xi_{x}^{iy,n}.x',\xi_{x}^{iy,n}.y'}-E_{\xi_{x}^{iy,n}.x',\xi_{x}^{iy,n}.y'}\right)\cdot 
P_{x}\xi_{x}^{iy,n}.w = 0
  \label{eq curved optimized derivative over L}
\end{equation}

Substituting Equations~\ref{eq curved optimized model} to \ref{eq curved optimized derivative over p}
and Equation~\ref{eq curved optimized derivative over L}, and moving the part with the measured detector pixel
counts to the right-hand side we get the final form of the system of equations for $P_x$ and $L_{iy}$:
\begin{eqnarray}
\sum_{iy,n} \sum_m P_{\zeta_{\xi_{x}^{iy,n}.x',\xi_{x}^{iy,n}.y'}^m.x'}\cdot
L_{\zeta_{\xi_{x}^{iy,n}.x',\xi_{x}^{iy,n}.y'}^m.iy'}
{\zeta_{\xi_{x}^{iy,n}.x',\xi_{x}^{iy,n}.y'}^m.w}\cdot \nonumber \\
\cdot L_{iy}\cdot\xi_{x}^{iy,n}.w =
\sum_{iy,n}E_{\xi_{x}^{iy,n}.x',\xi_{x}^{iy,n}.y'}\cdot L_{iy}\cdot\xi_{x}^{iy,n}.w & & \label{eq curved equation for P}\\
\sum_{x,n} \sum_m P_{\zeta_{\xi_{x}^{iy,n}.x',\xi_{x}^{iy,n}.y'}^m.x'}\cdot
L_{\zeta_{\xi_{x}^{iy,n}.x',\xi_{x}^{iy,n}.y'}^m.iy'}\cdot
{\zeta_{\xi_{x}^{iy,n}.x',\xi_{x}^{iy,n}.y'}^m.w}\cdot \nonumber \\
\cdot P_{x}\cdot\xi_{x}^{iy,n}.w =
\sum_{x,n} E_{\xi_{x}^{iy,n}.x',\xi_{x}^{iy,n}.y'}\cdot P_{x}\cdot\xi_{x}^{iy,n}.w 
\label{eq curved equation for L}
\end{eqnarray}

The software implementation for Equations~\ref{eq curved equation for P} and \ref{eq curved equation for L}
faces two challenges: the construction of $\xi$ and $\zeta$ tensors and the construction of the matrices and RHS's. The first
is solved through a single loop over all subpixels of $L$ for each column $x$. In this loop one can record detector pixel
coordinates and the corresponding footprints, which are what $\xi$ stores. In the same loop for each detector pixel one
records the coordinates of the contributing subpixel and its footprint filling the $\zeta$ tensor.

The second challenge comes from the fact that the indexing of the unknown vectors ($P$ and $L$) in equations
\ref{eq curved equation for P} and \ref{eq curved equation for L} is not sequential. One should regard
this as a permutation of rows and columns in the linear systems of equations ${A}\cdot x = y$.
The equation permutation needs to be stored in order to recover the correct order of elements in the unknown vectors.

Finally, at the horizontal ends of the swath in some rows the slit image can (due to the tilt) stretch outside the data fragment as schematically shown in \autoref{tilted slit}.
The simplest way to handle this issue is to pad each swath with additional columns on both sides and then clip
the extracted spectrum by the corresponding amount. At the edges of the detector padding is not possible, which may require the clipping of the extracted spectrum.

\begin{figure*}[hp!]
\begin{center}
\includegraphics[width=1.464\columnwidth]{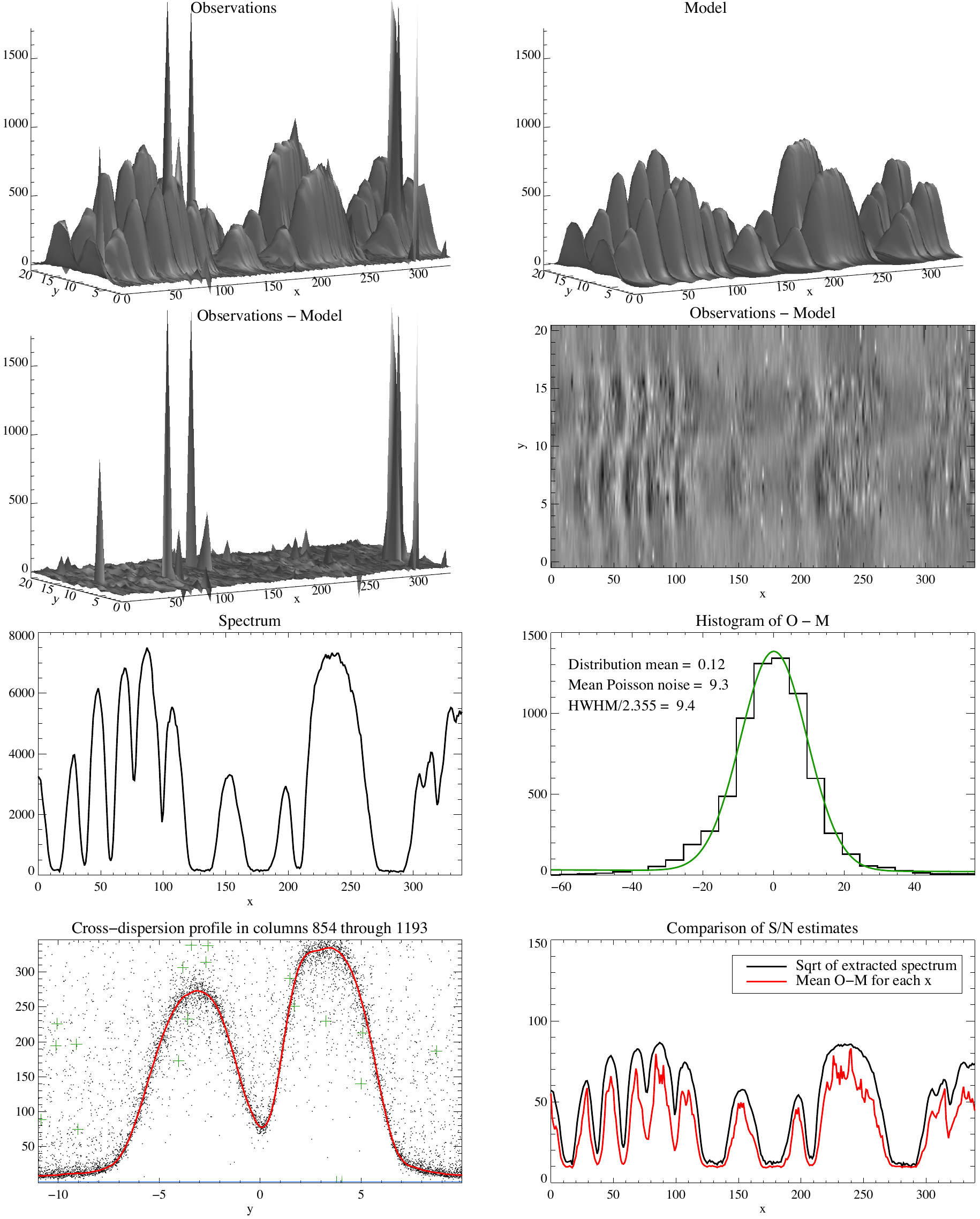}
\includegraphics[width=1.464\columnwidth]{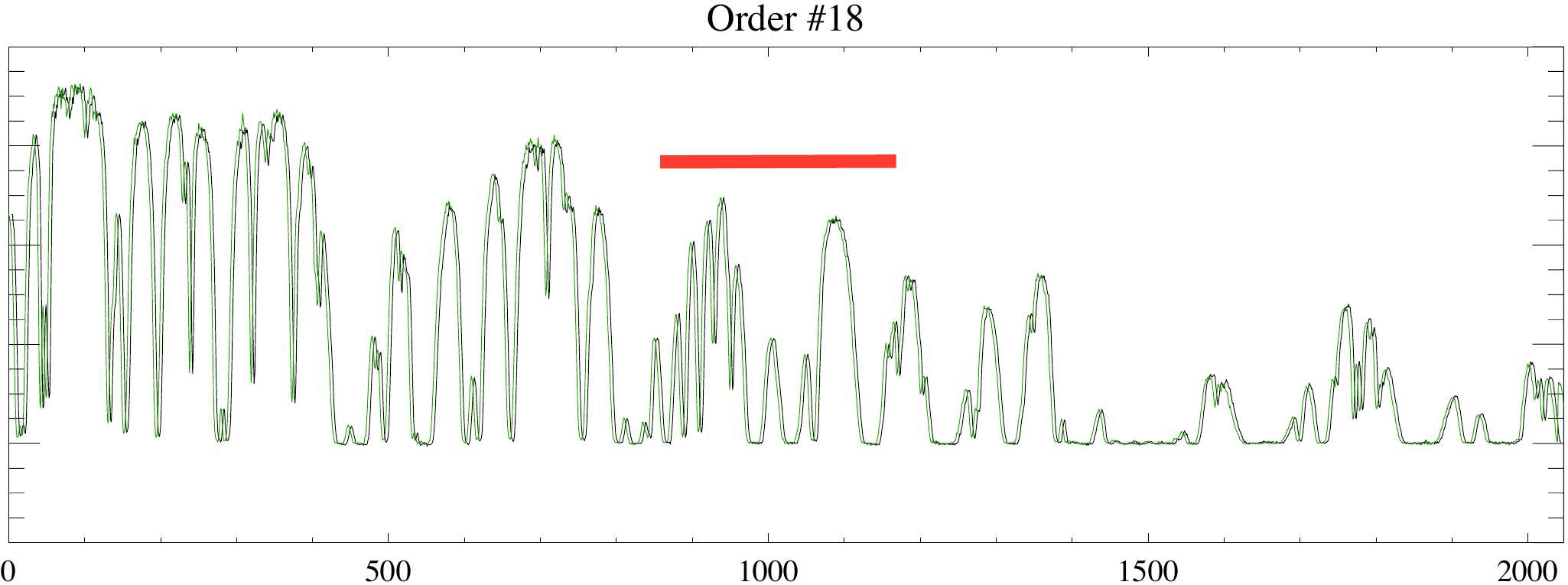}
\caption{Slit decomposition and uncertainty estimates based on data - model statistics. The top four panels show
the 2D image of the data as registered by CARMENES ($E_{x,y}$ in equations), the model reconstructed using the deduced $P$ and $L$ functions ($S_{x,y}$) and their difference. The difference image on the right
of the 2\textsuperscript{nd} row was
multiplied by the bad pixel mask constructed during the decomposition and shows residual ripples with an amplitude
of $\sim$12.5 counts.} The next four panels show the recovered spectrum ($P$) and slit illumination function ($L$) on
the left. Scattered dots are the actual data points aligned with the order center and divided by the extracted
spectrum. Green pluses show rejected (masked) pixels. The right panels show the comparison of the uncertainty estimate with a Poissonian noise estimate for the whole swath (histogram) as well as for each column. The bottom panel compares our extraction (in green) with the standard CARMENES pipeline (in black). The black line was shifted to the right by 1 pixel for visibility.
\label{extraction uncertainties}
\end{center}
\end{figure*}

\subsection{Uncertainties of extracted spectra}

{
The extraction procedure described above can be seen as an inverse problem of a convolution type. It helps avoiding complications due to the degeneracy between $P$ and $L$ or noise amplifications in areas of low signal. Error
propagation is notoriously difficult for inverse problems, since the measured noise is known for the result of the
convolution i.e. data, while the model statistics is unknown \textit{a priori} as is the transformation from detector
pixels to the $P,L$ space. Thus, we take a different approach. Once we have a converged model for a given swath,
we can construct the distribution of the difference between observations and model for the whole swath and for each "slit image"
realisation, indexed by column number $x$. The full swath distribution is obviously better defined, so we fit a
Gaussian to it. The standard deviation for this Gaussian can be compared to the Poisson estimate using the extracted
spectrum $P$.

To illustrate the procedure we use a challenge suggested by the referee and used a CARMENES \citep{2016SPIE.9908E..12Q} near-IR
spectrum of HD209458 (car-20180905T23h01m44s-sci-czes-nir). We selected order 18, which has some columns with high
signal as well as some with no apparent stellar continuum due to the strong water absorption in the Earth atmosphere.
\autoref{extraction uncertainties} shows the application of vertical slit decomposition and how uncertainties of the extracted
spectrum are estimated. We use columns 854-1193 of order 18 (0th order at the bottom) as an example of the first
detector in the near-IR arm of CARMENES. The standard deviation estimated for the whole swath using the histogram of the
data-model differences (panel in the middle right) is nearly identical to the mean Poisson statistics estimate
$\sum_{x,y} \sqrt{E_{x,y}}/n_x/n_y$. The panel below uses a similar approach for individual slit images as indexed by
the column number. The plot shows two different estimates for the signal-to-noise ratio. The black line is
the square root of the extracted spectrum, i.e. a simple Poisson distribution, while the mean deviation between the
data and the model weighted by pixel contribution to the given slit image is plotted in red. The noticeably higher
level of the uncertainty estimate from the slit decomposition (lower S/N) is actually real. It reflects the
shortcut we took in this test extraction by ignoring the effective "tilt" of the slit image created by the image
slicer (two half-circle images of the input fiber), which is well seen in differences (right panel in the
2\textsuperscript{nd} row). While the amplitude of the difference is small, it still drives up our
uncertainty estimates. The impact on the extracted spectrum is negligible as illustrated by the bottom panel of
\autoref{extraction uncertainties}, where we compare our extraction with the standard CARMENES pipeline output.
The two extracted spectra agree to 0.1\%, if we ignore a few "bad pixels" present in this swath.
}

\section{Curvature Determination}
\label{sec:curvature}
\begin{figure}
    \centering
    \includegraphics[width=0.7\columnwidth]{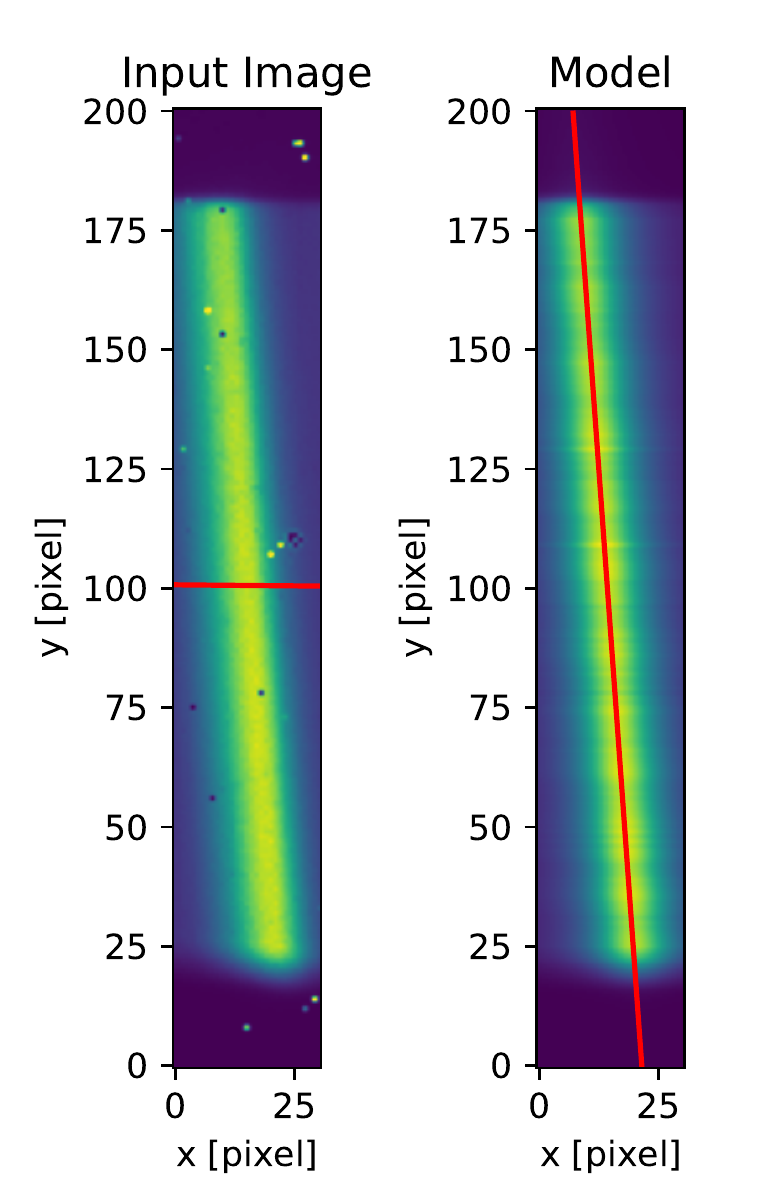}
    \caption{Curvature estimation for one spectral line. Left: The rectified input image cut out from the
    wavelength calibration frame. The order trace $y_c(x)$ is marked in red. Right: The best fit
    model image. The extracted tilt is marked in red. Note that we only fit a linear tilt here.
    Data from ESO/CRIRES+ Fabry-P\'erot interferometer.}
    \label{fig:shear_estimate}
\end{figure}

The new "curved slit" extraction algorithm presented in \autoref{sec:method} can account
for the curvature of the slit on the detector, but assumes that the shape of the slit image is known
\textit{a priori} at any position on the science detector. The tilt and the curvature of the slit
image are usually the result of compromises made when selecting the optical scheme of a
spectrometer and detector orientation in the focal plane. These may lead to a significant
average tilt, but in general the slit image shape will vary slowly along the dispersion
direction and between spectral orders. The curvature of the slit is always small and
hardly important for the observation of point sources. However, it may introduce a shift of the
wavelength scale, if a different part of the slit is used for the wavelength calibration.
Assuming a slow change of the slit image, we can measure the shape at a few places in each
spectral order and then interpolate to all columns. This can be done using e.g. the wavelength
calibration (\autoref{sec:wavecal}) data following the steps outlined below. An important
prerequisite for this to work is an existing order tracing that provides a center line for
each order. The center line follows the image of a selected reference slit point (e.g. the middle of
the slit spatial extension) across the whole spectral format. In a given spectral order it
is a function $y_c(x)$ relating column number $x$ with the vertical position
of the trace line $y$. We postulate that for an integer value of $x$ the center of the slit image
in dispersion direction falls precisely onto the middle of the pixel column $x$.
Tracing the slit image up or down from the reference position the center of the slit $y$ may shift
left or right from the center of column $x$ due to the tilt and curvature of the image.

\begin{figure}
    \centering
    \includegraphics[width=1.\columnwidth]{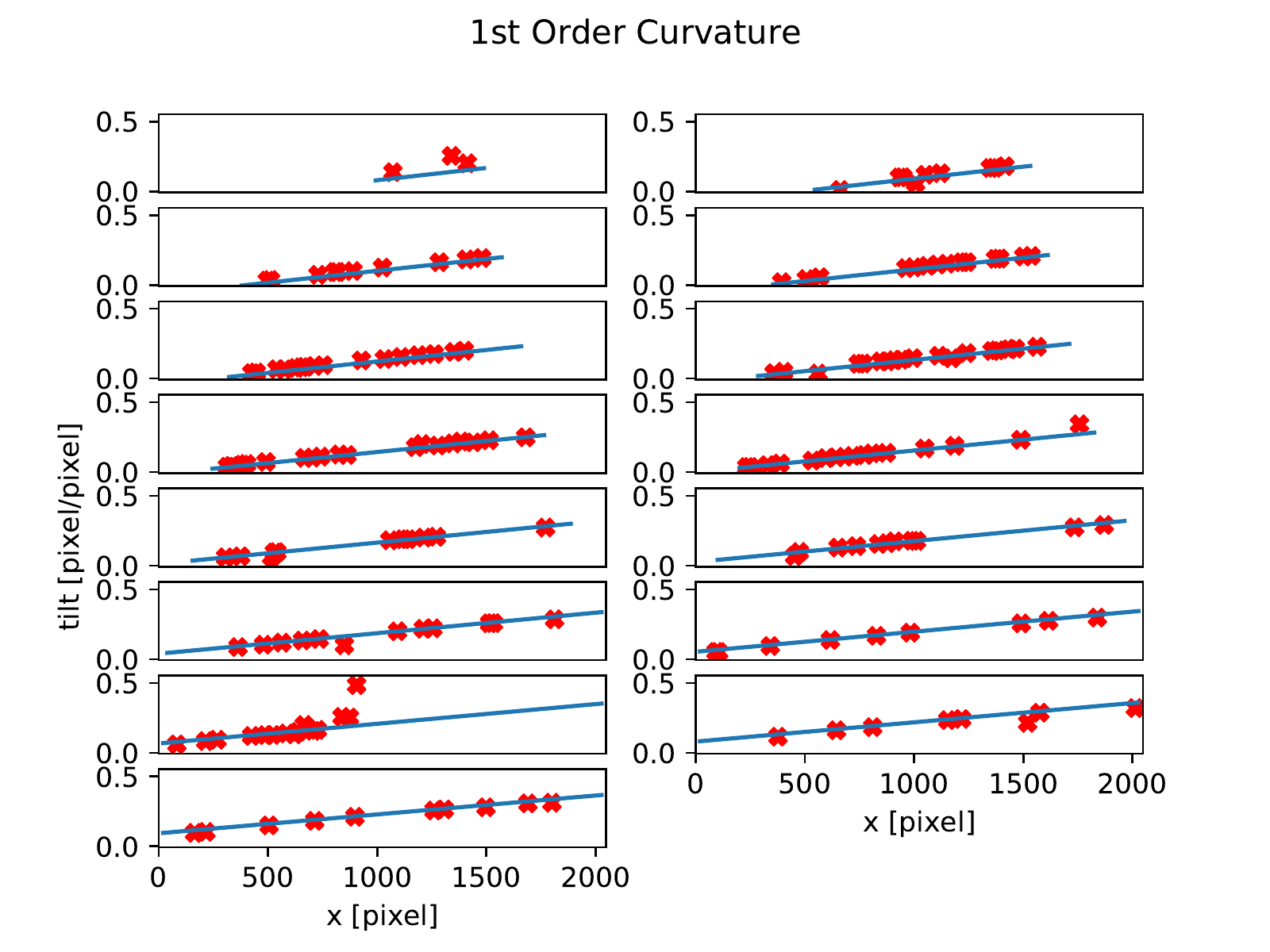}
    \caption{Tilt variations per order (each panel is one order). Red pluses show
    the tilt determined for each individual emission line. The blue line is the 2D polynomial
    fit through all lines in all orders. Data from VLT/X-Shooter \citep{2011A&A...536A.105V}.}
    \label{fig:shear_curves}
\end{figure}

\begin{figure}
    \centering
    \includegraphics[width=1.\columnwidth]{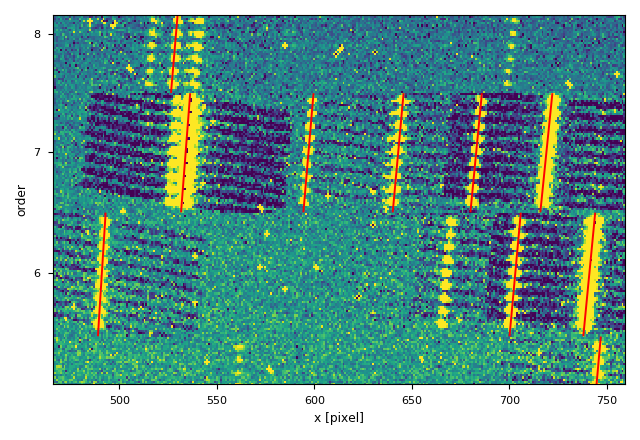}
    \caption{A fragment of the input image used for the slit curvature determination (blue-yellow),
    with the recovered curvature (red) at line positions plotted on top. The red line is constructed
    from the final fit of the curvature and tracks the center of each spectral line along the slit.
    Data from VLT/X-Shooter \citep{2011A&A...536A.105V}.}
    \label{fig:shear_final}
\end{figure}

We model the slit image shape using the wavelength calibration in the following three-step procedure: 
\begin{enumerate}
    \item We identify emission lines in the wavelength calibration spectrum and select the "good" lines based on their intensity (not too faint by comparison with the noise, not saturated and not blended).

    \item For each selected line $i$ we fit a 2D-model to the line image. The model consists of a Gaussian
    (or Lorentzian) in dispersion direction with three parameters (line position,  line strength, and
    line width). Due to the tilt and curvature of the slit image the line position may shift along the
    row (left or right, $\delta x$) as we move away (up or down, $\delta y$) from the reference position
    given by the order trace $y_c(x)$ as shown in \autoref{fig:shear_estimate}. The offset $\delta x$ is
    given as a function of the vertical distance from the central line: 
    $\delta x = a_i\ \delta y(x)^2 + b_i\ \delta y(x)$, where $\delta y(x) = y - y_c(x)$. Note, that
    when $y$ is equal $y_c(x)$, $\delta x$ is zero by
    definition, so there are only two coefficients to fit corresponding to tilt and curvature. When
    the tilt is large, the width of the line may be overestimated, but that does not affect the center
    position of the curve. To account for the slit illumination function, and to avoid problems with
    fitting the amplitude of the model for each row individually, we simply scale the fit by the
    median of the data in each row.

    \item Finally, we combine all coefficients derived for individual emission lines by fitting tilt 
    and curvature variation across all the orders as a function of the order column and order number.
    The curvature at each position is then described by:

    \begin{eqnarray}
        \mathrm{curvature:}\    a(x, m) &=& c_1(x, m) \ \delta y + c_2(x, m) \ \delta y^2 \nonumber \\
        \mathrm{tilt:}\    b(x, m) &=& d_1(x, m) \ \delta y + d_2(x, m) \ \delta y^2 \label{eq:curvature}
    \end{eqnarray}

    where $x$ is column number, $\delta y$ is the row $y$ distance to the central line, and $m$ is the
    order number. \autoref{fig:shear_curves} shows an example of the fitted tilt. Note that this is a 2D
    polynomial fit, so we reconstruct the slit image shape even in parts of spectral orders without any
    emission lines in the input image. This step is crucial for detecting and removing outliers created
    by a failed fit to individual lines (e.g. due to a cosmic ray hit or a detector defect). Similar to
    the wavelength calibration, the choice of the degree of the fit should be adjusted depending on the
    instrument and density of useful emission lines.
    The same curvature is plotted on top of the input image in \autoref{fig:shear_final}.
\end{enumerate}

\section{Wavelength Calibration}
\label{sec:wavecal}
The wavelength calibration of a grating spectrometer is based on the grating equation that connects
the wavelength $\lambda$, the physical spectral order number $m$, the spacing between grooves
$\delta$, and the dispersion angle $\beta$:
\begin{equation}
    \lambda = \delta(\sin\alpha + \sin\beta)/m
    \label{eq grating}
\end{equation}
where $\alpha$ is the incidence angle, independent of the wavelength. The angular dispersion
$d\lambda/d\beta$ is a function of $m$ and and reflection angle $\beta$ ($\cos\beta$). For
modern echelle spectrometers $\beta$ occupies a small range of values centered on the blaze
angle. The latter has typical values between 60 and 80 degrees. These result in a nearly
constant dispersion for any given order. Thus, the relation between pixels and their wavelengths
can be represented by a low-order polynomial. Polynomial orders of 3 to 5 are usually sufficient
to reproduce the $\lambda - \beta$ relation and to catch possible distortions introduced by the
imaging system.

The determination of the polynomial coefficients requires a reference source with precision
wavelengths assigned to emission (or absorption, as in the case of an absorption gas cell) lines.
In this paper we leave out such crucial steps of wavelength calibration as the determination of line
centers and the use of a laser frequency comb (LFC) or Fabry-P\'erot etalon calibration source. We plan
to re-visit these aspects in a separate paper. In the meantime one can find a detailed description
of the calibration processing in the paper by \cite{2020MNRAS.493.3997M}.

Instead we summarise the procedure. A spectrum of a calibration source must be recorded with the spectrograph, spectral lines identified, their position measured in the detector coordinate system (pixels), and coefficients of polynomial regression determined. The spectral features of the reference source should preferably be evenly distributed across the spectral order to provide a homogeneous approximation and to minimize the maximum error. Note, that in this procedure the main uncertainties are frequently coming from the wrong identification of lines, measurements of their positions and the use of blended or saturated lines.

It is a standard practice to create a specific reference line list for each instrument and setting,
which includes the expected positions on the detector as well as the laboratory wavelength for each
line. Once the solution is obtained, it can be recycled for later wavelength calibrations assuming
that any line position changes will be much smaller than the line separation in dispersion direction and
less than the order separation in cross-dispersion direction. An existing solution is then used as an
initial guess for the next wavelength calibration.

\begin{figure}
        \centering
        \includegraphics[width=1.\columnwidth]{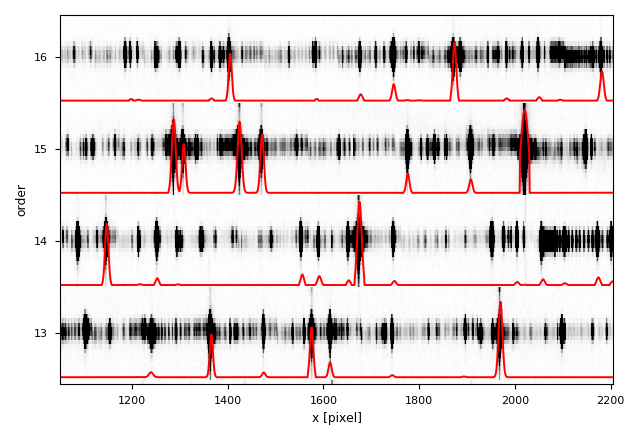}
        \caption{A fragment with four partial orders of the input image used for the wavelength
        calibration (greyscale). The reference spectra (red) have been shifted and scaled along the y axis so that they
        overlay their respective orders. Data from La Silla/HARPS \citep{2003Msngr.114...20M}.}
        \label{fig:wave1}
    \end{figure}

The observed wavelength calibration spectrum is often extracted with a simple summation across
the order. This is often sufficient since the flux level of a lamp tends to be much higher than
that of a star. An example of the extracted image is shown in \autoref{fig:wave1}.

For many applications, instruments like HARPS \citep{2003Msngr.114...20M}, which are designed for high-stability, with no moving parts, and located in stabilized environment, may use calibrations taken several hours before or after the science data. For extreme precision measurements as well as for general purpose instruments, the required repeatability is not reached this way. In these cases, an attached or simultaneous calibration is needed to complement the science data. Here again the reference solution can be used as an initial guess.

Finally the best fit polynomial connecting position and wavelength can be determined.
\cite{1994PhDT........16V} proposed to use a 2D polynomial matching the dispersion variation
within each order and between orders as expected from the grating equation~\ref{eq grating}.
The requirement of smooth variation (low order polynomial) of dispersion and central wavelengths
between spectral orders sets additional restrictions on the solution and helps constraining the
polynomial in parts of the focal plane void of emission lines in the calibration spectrum.
Additional discussion of the polynomial degree and the dimensionality can be found in
\autoref{ssec:wave_degree}.

The polynomial fit involves a gradual improvement of the solution by rejecting the largest outliers in an iterative process. For this purpose the residual $R_i$ is defined not in the wavelength, but in velocity space:
\begin{equation}
    R_i = \frac{\lambda_\text{ref} - \lambda_\text{obs}}{\lambda_\text{ref} } c_\text{light}. \label{eq:wavecal_residual}
\end{equation}
The process follows the conventional sigma-clipping algorithm and stops when no more statistically-improbable outliers are found.

Starting from a reference solution and applying the outlier rejection described above, some lines may be unidentified. They can be recovered in an auto-identification phase. It finds all suitable unidentified peaks in the calibration, estimates their measured wavelength using the reference solution, and searches the reference lamp atlas for a possible match. "Suitable" lines are defined after Gaussian fitting as having a FWHM in an acceptable range for the given instrument. The atlas is also checked for any possible blending of lines. Sigma-clipping and auto-identification phases can be repeated more than once to ensure the convergence of the overall procedure.

\subsection{2D versus 1D Wavelength Polynomial}
    \label{ssec:wave_degree}
\begin{figure*}
        \centering
        \includegraphics[width=0.99\columnwidth]{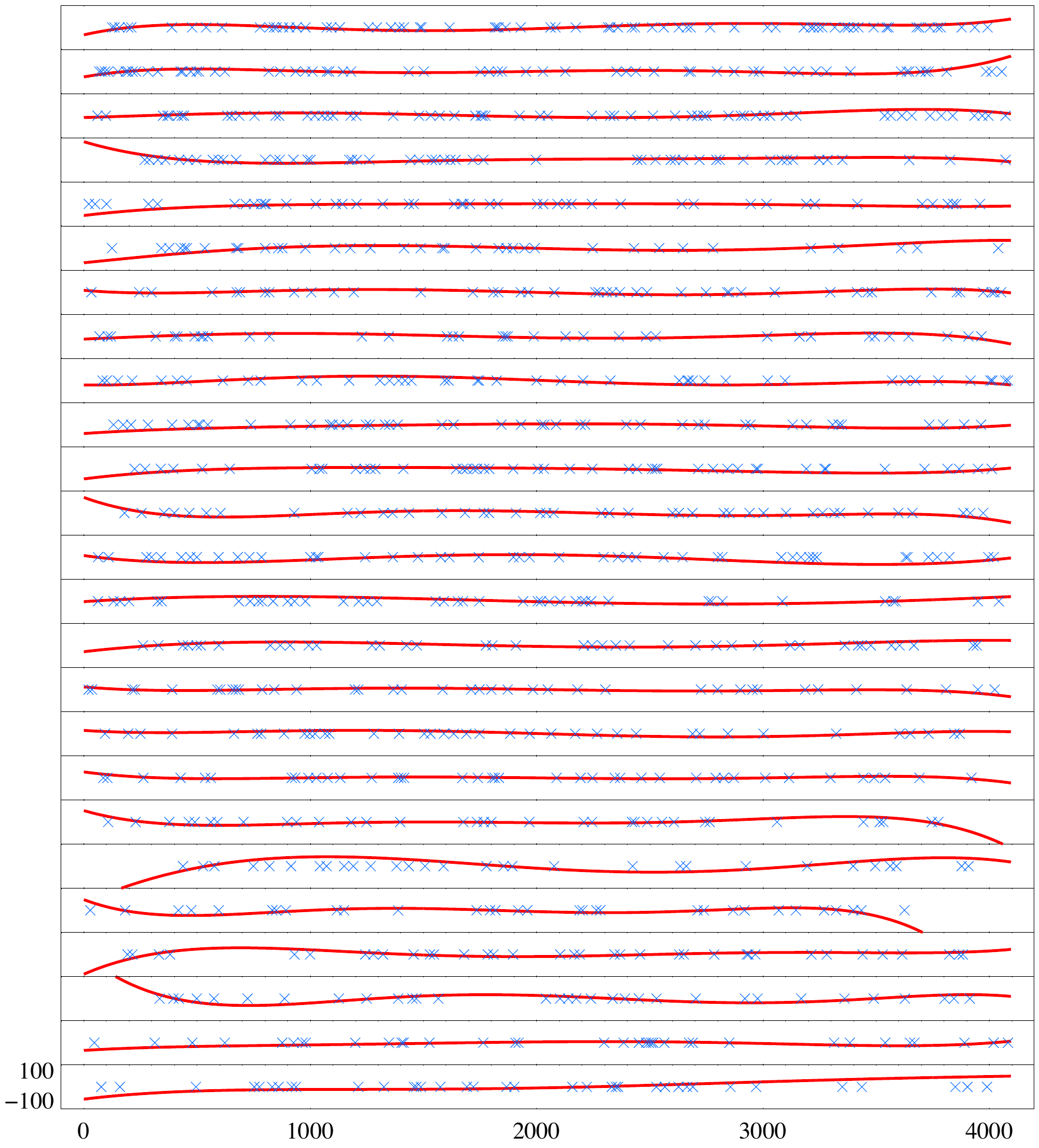}
        \includegraphics[width=0.99\columnwidth]{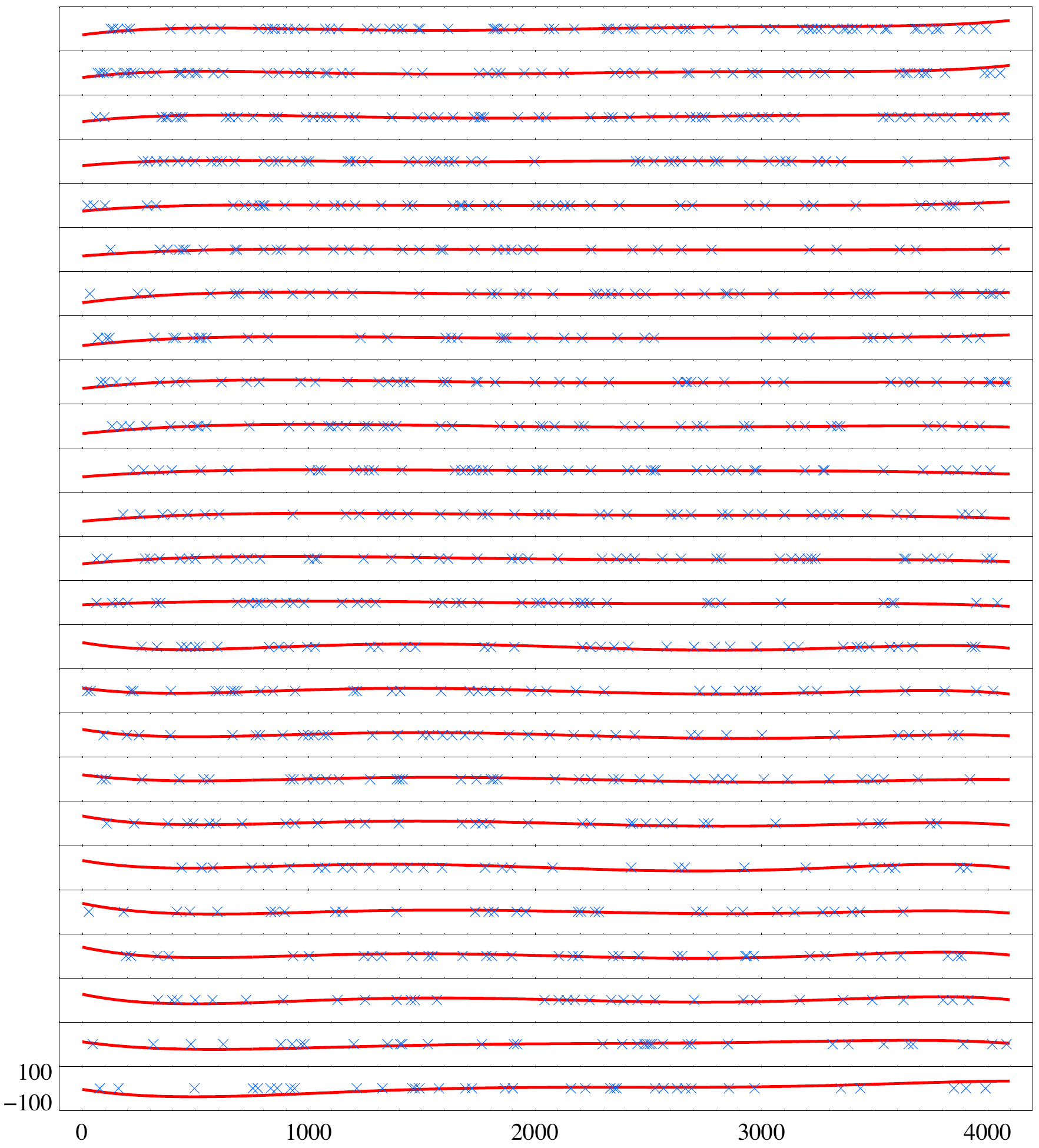}
        \caption{\sout{Relative} Difference between the LFC wavelength solution and the ThAr-based 1D (left) and 2D (right) solutions for the red detector arm of the La Silla/HARPS spectrometer. Each vertical panel shows one spectral
        order with the longest wavelength order positioned at the bottom. The differences between ThAr and
        LFC solutions in m/s are plotted as red lines against detector pixels in the X-axis. LFC solutions were
        constructed separately for each spectral order. Blue crosses indicate the positions of the ThAr lines used
        for the construction of the wavelength solutions. In all cases, a 5\textsuperscript{th} order polynomial
        was used for the dispersion direction. The 2D solution included an additional 3\textsuperscript{rd} order cross-dispersion component as well as the corresponding cross-terms.}
        \label{fig:1d_2d_wavecal}
    \end{figure*}

    A reoccurring discussion when performing wavelength calibration is whether to use one 2D polynomial for all orders, or to use individual 1D polynomials for each order. The main arguments revolve around the ability of a 2D solution to minimize the maximum error in parts of the spectral orders without any reference lines versus the flexibility of the individual polynomial fit to each order as illustrated in \autoref{fig:1d_2d_wavecal}.
    In this example, based on the ThAr and LFC wavelength calibrations of the La Silla HARPS spectrometer, the differences between the 1D and the LFC solution reach in excess of 200 m/s in the peripheral parts of the detector where a non-homogeneous distribution of the calibration lines is aggravated by the low signal. A similar 2D ThAr solution is generally close to the LFC result, with the exception of the very ends of spectral orders.

    This question is also closely related to the degree of polynomials used for the fit. A higher order polynomial
    can fit the data better, but may also have larger variations from the true solution in places where data
    points are sparse. Ideally one wants to determine the best representation of the data, with the fewest parameters possible. This is where the Akaike information criterion (AIC) \citep{1974ITAC...19..716A} is useful, as
    it combines the goodness of fit and the number of parameters into a single measure that can easily be
    compared between solutions. The AIC is defined as:
    \begin{equation}
        \text{AIC} = 2 k - 2 \ln{L}, \label{eq:aic}
    \end{equation}
    where k is the number of parameters in the model and $L$ is the likelihood. Since the least
    squares fit was used for the model the likelihood is given by the squared sum of the residuals
    (also known as the $\chi^2$):
    \begin{equation}
        \ln{L} = - \frac{N}{2} \ln\left[\sum_{i=0}^N \left(R_i / c_{light}\right)^2\right] + C, \label{eq:likelihood}
    \end{equation}
    where $N$ is the number of lines, $R_i$ is the residual as defined by \autoref{eq:wavecal_residual}, and $C$ is a constant factor that we can ignore, since only the difference between AICs is relevant. Similarly the $c_{light}$ factor could be removed since it only results in a constant, but is kept to make the values dimensionless.
    
\begin{figure}
        \centering
        \includegraphics[width=1.\columnwidth]{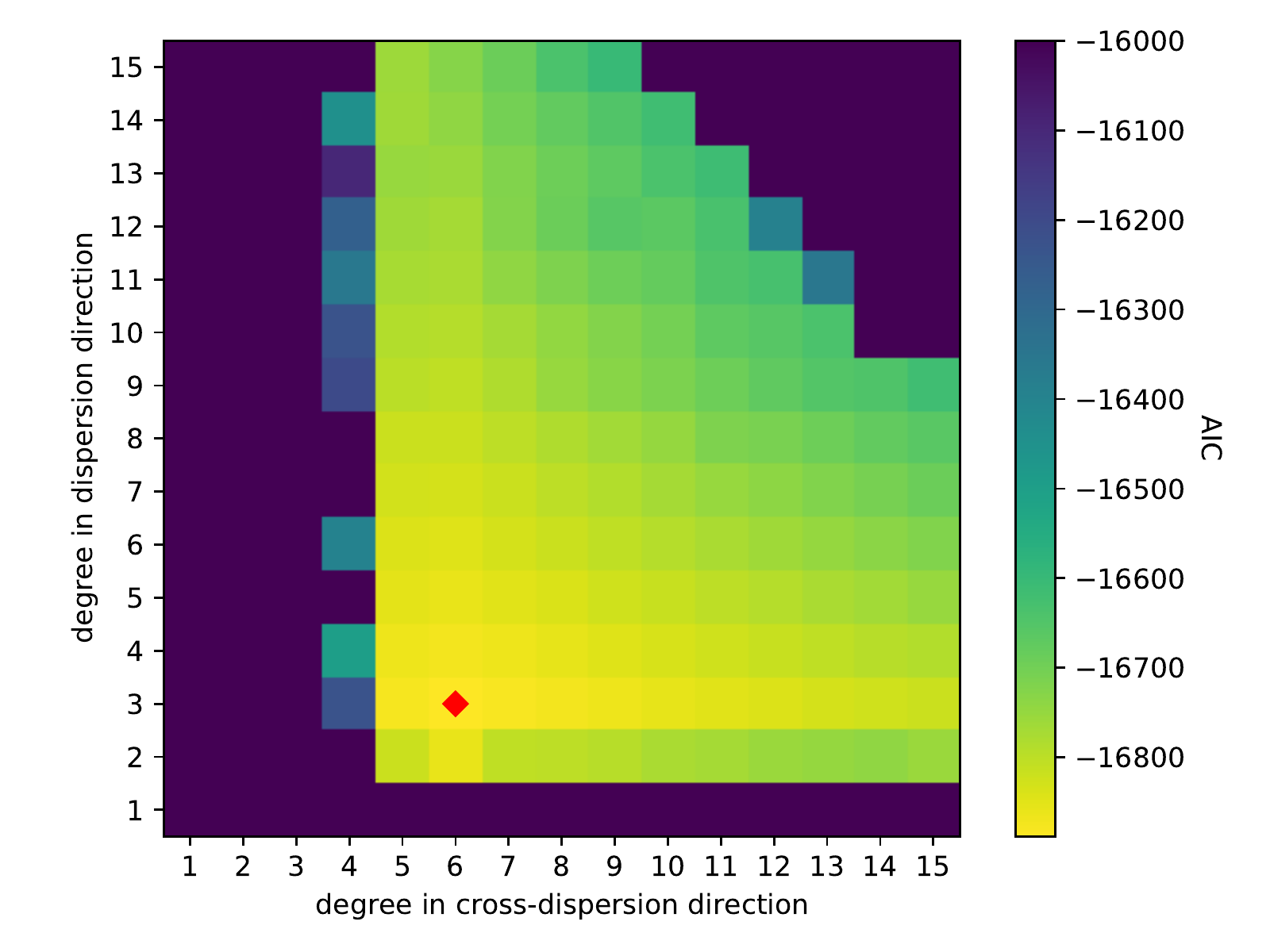}
        \caption{The AIC of different polynomials to the wavelength calibration with a ThAr gas lamp, based
        on the degree of the polynomial. The red diamond marks the best AIC(lower is better). Note that all values above
        -16,000 are shown in the same colour to make the gradient more visible. Data from La Silla/HARPS.}
        \label{fig:wavecal_aic_grid}
    \end{figure}

    Then we can simply use a grid search to find the best model, i.e. the one with the lowest AIC. For the example of the HARPS ThAr wavelength calibration, the best AIC value is achieved with a 2D fit with degrees 3 and 6 in dispersion and spatial direction respectively, see also \autoref{fig:wavecal_aic_grid} for an overview of the parameter space. The best 1D fit for this example is achieved with a polynomial of degree 2, although the AIC is larger than that of the 2D fit. 
    Using HARPS with an LFC reveals the detector stitching as discussed by \citet{2019A&A...629A..27C}. We can include corrections in the fit and find that the best fit has the degrees 9 and 7, respectively. Notably the spatial degree remains similar, as the order number is separate from the detector pixels. As for ThAr the 2D fit is preferred over the 1D fit.
    
\begin{figure}
        \centering
        \includegraphics[width=1.\columnwidth]{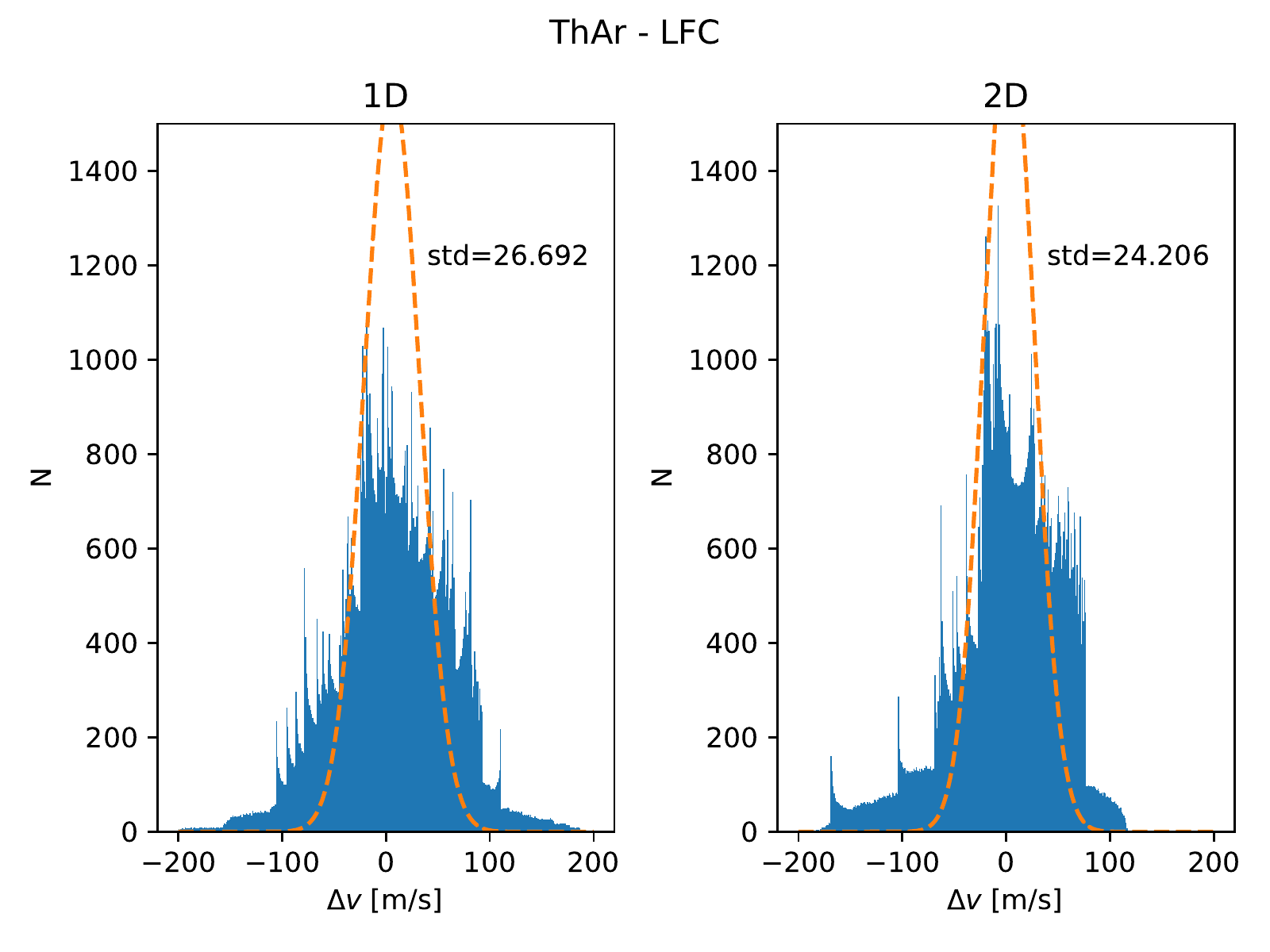}
        \caption{Distribution of the difference between the ThAr wavelength solution and the LFC wavelength
        solution, evaluated for each point in all orders. The LFC solution is based on a 2D polynomial.
        Note that there are more outliers beyond the limits of the histogram, which are not shown here.
        The orange dashed line, shows the Gaussian with the same mean and standard deviation as the whole
        distribution. Left: 1D ThAr solution, Right: 2D ThAr solution.}
        \label{fig:wavecal_residual_distribution}
    \end{figure}

\begin{figure}
        \centering
        \includegraphics[width=1.\columnwidth]{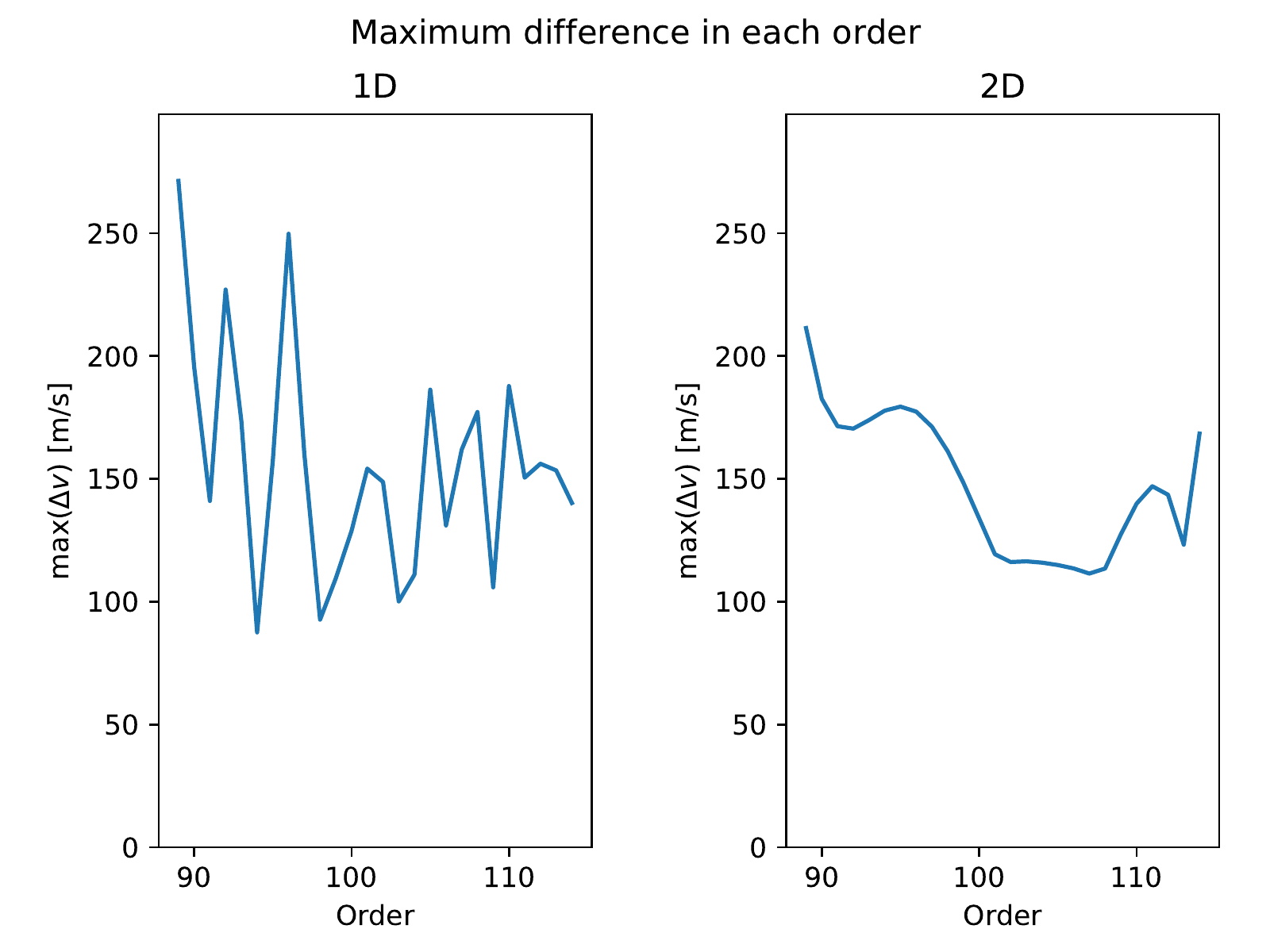}
        \caption{Maximum difference between the ThAr wavelength solution and the LFC wavelength solution in each order. Left: 1D ThAr solution, Right: 2D ThAr solution. The data from the red CCD of HARPS at La Silla is showing spectral orders between 89 and 114.}
        \label{fig:wavecal_residual_maximum}
    \end{figure}

    We can also compare the results of the different models with the LFC solution as an alternative reference. \autoref{fig:wavecal_residual_distribution} shows the distribution of the differences between the ThAr solutions and the LFC solution for HARPS. The two distributions are on the same order of magnitude, with the 1D solution being slightly wider. Notably the 1D solution has more outliers, as shown by the larger standard deviation of the Gaussian. This is also visible in \autoref{fig:wavecal_residual_maximum}, as here the largest difference in each order is clearly larger in the 1D solution, compared to the 2D solution.

    This is exactly what we expected and we conclude that at least for the ThAr  calibration the 2D solution
    is more robust against missing data and possibly against errors in the line center measurements.
    
\section{Continuum Normalization}
\label{sec:continuum}
    Robust continuum normalization is a notoriously difficult task, even for well-behaved absorption line spectra. The few exceptions include cases when there is a well-matching synthetic spectrum available (as in the case of solar flux), or hot stars with very few spectral features.
    Among the various attempts of attacking this problem better success was achieved with iterative schemes that fit the spectrum with a smooth function and gradually excluding points below the curve, until the distribution of data offsets from the constructed envelope becomes approximately symmetric and Gaussian. This, of course, does not guarantee that the normalized observed spectrum will match the synthetic spectrum. On the other hand, the correct synthetic spectrum is unknown to begin with and thus a heuristic approach is well motivated.
    The problem then becomes how to decide if a given point belongs to a spectral line and thus should be dropped from the fit. A power spectrum analysis, to separate spatial frequencies associated with spectral
    lines from the continuum envelope, does not help when considering individual spectral orders one at a time.
    The continuum "diving" into the strong and broad lines remains just one of the issues.

    \subsection{Order splicing}
    \begin{figure}
        \centering
        \includegraphics[width=1.\columnwidth]{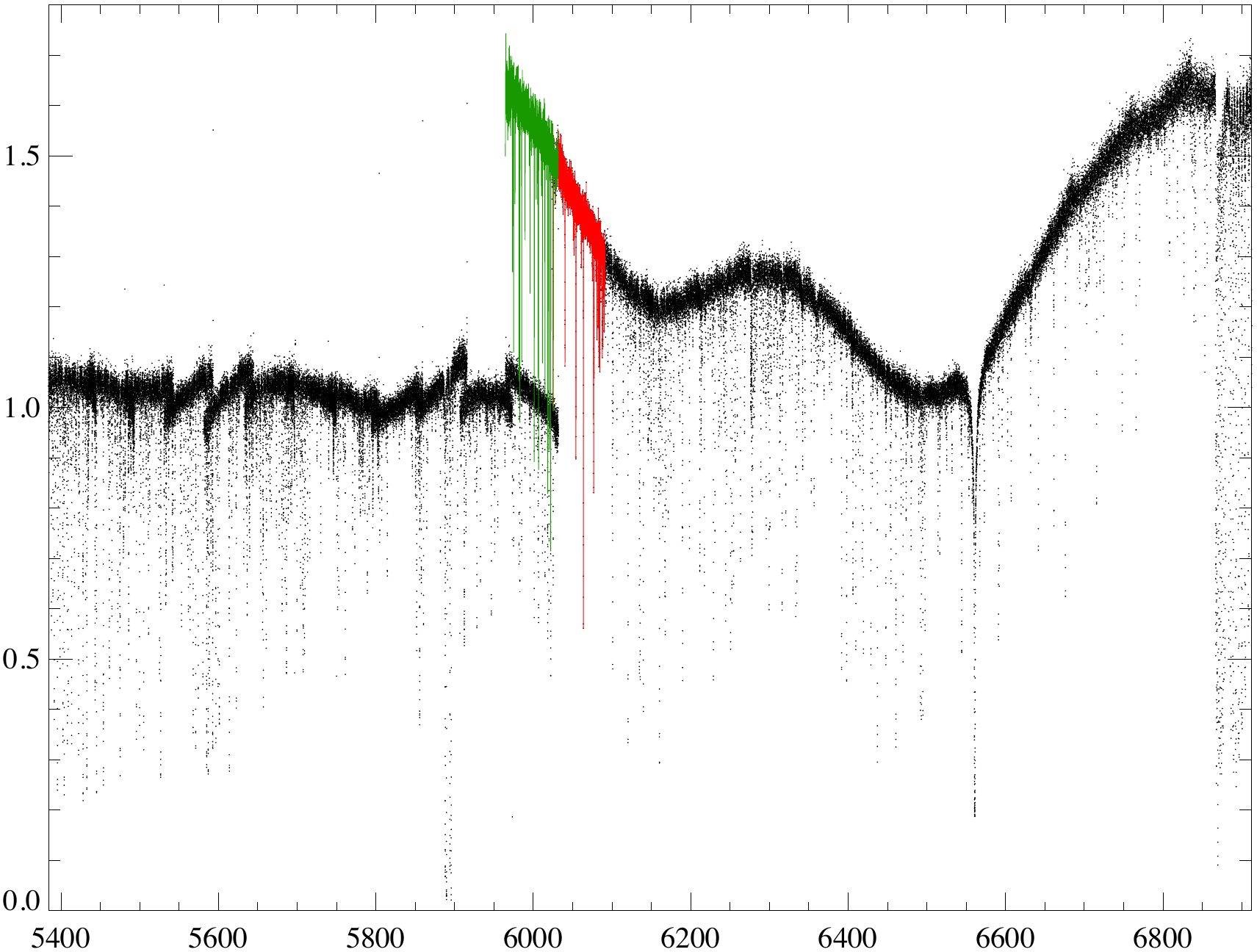}
        \caption{Example of the spectral order splicing using the common wavelength scale. Before splicing we
        divide the signal by the blaze function derived from the flat field. We start from the order with
        the highest S/N ratio, scale adjacent orders to achieve the best match in the overlap region and
        co-add the data with linear weights to avoid discontinuities. The black region to the right is already spliced. The scaling of the next order in line is
        shown in green. The scaled overlap region is shown
        in red. This example is from the red detector of La Silla/HARPS.}
        \label{fig:cont_splicing}
    \end{figure}

    In REDUCE we take a single-order approach to the next level by extending the range of the sampled spatial
    frequencies by splicing several spectral orders into a single long spectrum. Splicing requires an existing
    wavelength solution so that adjacent spectral orders can be aligned, scaled, interpolated, and co-added
    in the overlap region as illustrated in \autoref{fig:cont_splicing}. Combining the overlapping regions
    between neighbouring orders is complicated by the fact that the wavelengths associated with the pixels
    are different in the two orders. First we divide each order by the blaze function estimate,
    obtained e.g from the master flat field. Even though the blaze estimate is not a perfect continuum, it is a good first step towards flattening the individual orders. We then determine the wavelength overlap and  interpolate one
    order onto the wavelength grid of the other and vice versa. We finally co-add the values from
    both orders using a weighted sum with either linear weights or weights
    equal to the individual errors of each pixel\footnote{As the spectra in each order come from different
    pixels they are independent measurements.}. The sum is given by:
    \begin{equation}
        \Bar{s}_l(x) = \frac{x_r-x}{\sigma_l(x)\Delta x} \cdot s_l(x) + \frac{x-x_l}{\sigma_r(x)\Delta x}
        \cdot \Tilde{s}_r(x) \cdot
        \frac{\sigma_l(x) \cdot \sigma_r(x)}{\sigma_l(x) + \sigma_r(x)}
        \label{eq:splicing}
    \end{equation}
    where $x_l$, $x_r$ are the limits of the overlap region in pixels of the left order,
    $\Delta x = x_r - x_l$ and $\sigma$'s are the uncertainties.
    $\Bar{s}$ is the co-added value in the overlapping pixel $x$ of the left order indicated by subscript $l$.
    $\Tilde{s}_r$ is the overlapping part of the right order linearly interpolated onto pixel $x$ of the left
    order.

    The spliced spectrum still shows significant variations but they are rather smooth. "Waves"
    in the shape of the upper envelope are primarily coming from the spectrum of the flat field calibration
    (the source of the blaze functions) and spectral sensitivity of the detector. These variations are to be
    fitted in the following step, but they are described by much lower spatial frequencies than the spectral lines.
    Even H$\alpha$ looks "narrow" in comparison to the broad level variations in \autoref{fig:cont_splicing}.
    
    Uncertainties are spliced in the same fashion as the spectra to be used later in the fitting iterations.
    Once the splicing is completed we sort the wavelengths and interpolate the spectrum onto an equispaced
    wavelength grid to have a better handle on the frequency spectrum in preparation for the continuum fitting. 

\subsection{Continuum fitting}
    \begin{figure}
        \centering
        \includegraphics[width=1.\columnwidth]{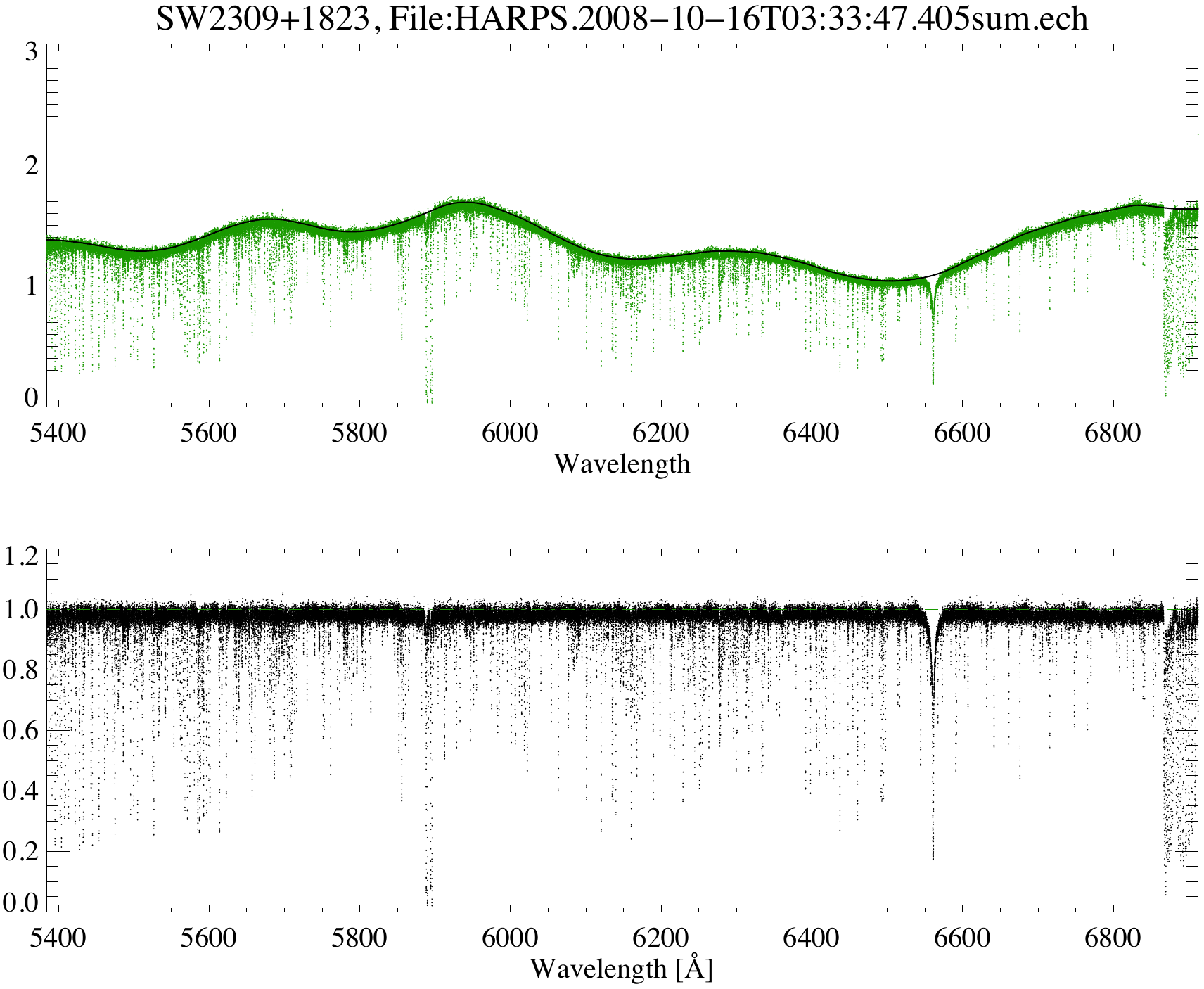}
        \includegraphics[width=1.\columnwidth]{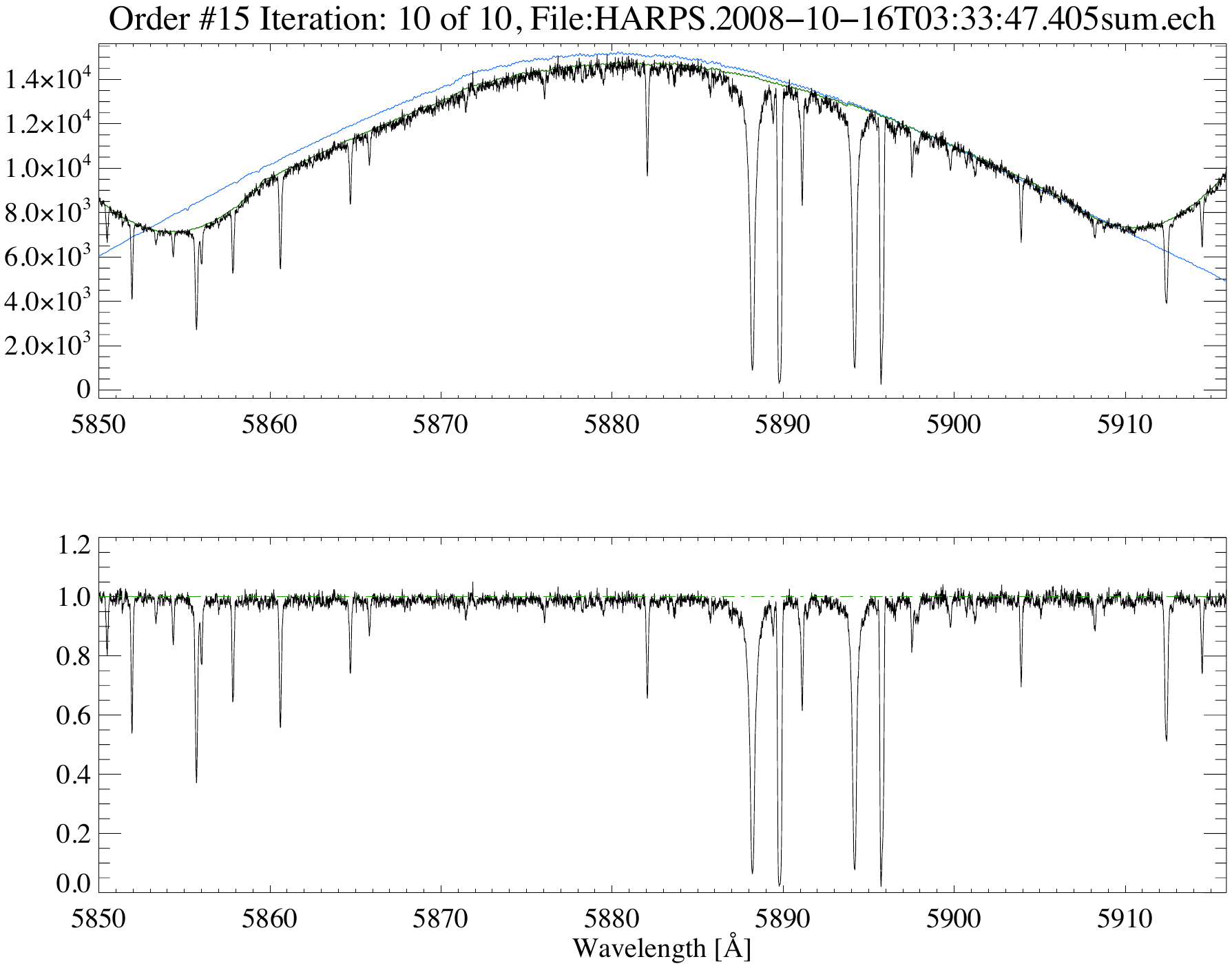}
        \caption{Final result of the continuum fitting.
        In the top, the spliced spectrum is shown in green while the continuum is in black. 2\textsuperscript{nd}
        panel: continuum normalized spectrum. Two bottom panels: zoom on order \#15 (absolute order 103). In the 3\textsuperscript{rd} panel from the top the spliced spectrum is in black, the original (non-spliced)
        blaze function is in blue and the continuum fit is in green. The continuum normalized order is shown in
        the bottom panel. Example is from the red detector of La Silla/HARPS.}
        \label{fig:cont_fitting}
    \end{figure}
    We use a custom-made filtering routine for the construction of a smooth non-analytical function.
    The fitting function $f(x)$ is defined in a such a way that it fits the data points well and at the same time
    has the least power in the highest spatial frequencies. The latter is achieved by restricting the
    minimization of the averages of the first and the second derivatives with two regularization terms:

    \begin{eqnarray}
        \sum_x \omega_x\left[f(x)-s(x)\right]^2+\Lambda_1\sum_x \left(\frac{df}{dx}\right)^2&+& \nonumber\\
                                               +\Lambda_2\sum_x \left(\frac{d^2f}{dx^2}\right)^2&=& \mathrm{min},
        \label{eq:opt_filter}
    \end{eqnarray}
    where $s$ is the spectrum, $x$ the wavelength point, $\omega$ is the weight/uncertainty, and $\Lambda_1$ and $\Lambda_2$ are regularization parameters that control the stiffness of the fit and its behavior at the end points. To be more specific, increasing $\Lambda_1$ makes the solution more horizontal while a larger $\Lambda_2$ ignores linear trends, but dumps high-frequency oscillations.
    The value of the two $\Lambda$ parameters needs to be adjusted empirically.
    From \autoref{eq:opt_filter} we construct a band-diagonal system of linear equations. Once the solution $f$ is obtained we can start the iterations by constructing the histogram of $s-f$ for all $s$ values that are larger than $f$ and estimating its width. We use a mirror reflection of the histogram in respect to 0 before fitting a Gaussian to it.
    The derived standard deviation is then used to reject points in $s$ that are well below $f$. The procedure is
    repeated with the remaining points starting from a recomputed $f$. We also verify the consistency between the
    (spliced) uncertainties and the standard deviation of the distribution. The process reaches convergence (no more points are rejected) in about 6-9 iterations. Examples of the final results are presented in \autoref{fig:cont_fitting}.
    After completing the continuum fit on an equispaced grid we interpolate the fit back to the initial wavelength grid of every spectral order. The same can be done with the observed spectrum, making
    individual orders look similar to the third panel from the top in Figure~\ref{fig:cont_fitting}.
    The gain is some increase in signal towards the ends of spectral orders. The downside is a possible
    loss of spectral resolution as well as distortions of the PSF due to focal plane aberrations and the
    interpolation procedures involved. Alternatively, one can convert the "spliced" continuum to a
    non-spliced version for each order using the splicing factors derived in \autoref{eq:splicing}.
    This way we do not modify the original data, which is important when e.g. the science goals include the accurate determination of radial velocities or analysis of spectral line profiles.

    We conclude by re-iterating that robust continuum normalization of observed spectra is \textit{impossible}.
    What is described above may or may not give a satisfactory solution depending on S/N, spectral line width
    (e.g. due to stellar rotation), quality of the blaze functions, and many other factors. A good selection of the stiffness parameters requires some experience. Finally, the spectral format and the overlap between spectral orders are crucial for the splicing and for the whole procedure we developed. For instruments that leave gaps in spectral coverage continuum normalization will remain an art, not science.

\section{Implementation}
\label{sec:implementation}
\subsection{\pyreduce}
       \begin{figure*}[ht]
        \centering
        \includegraphics[width=1.8\columnwidth]{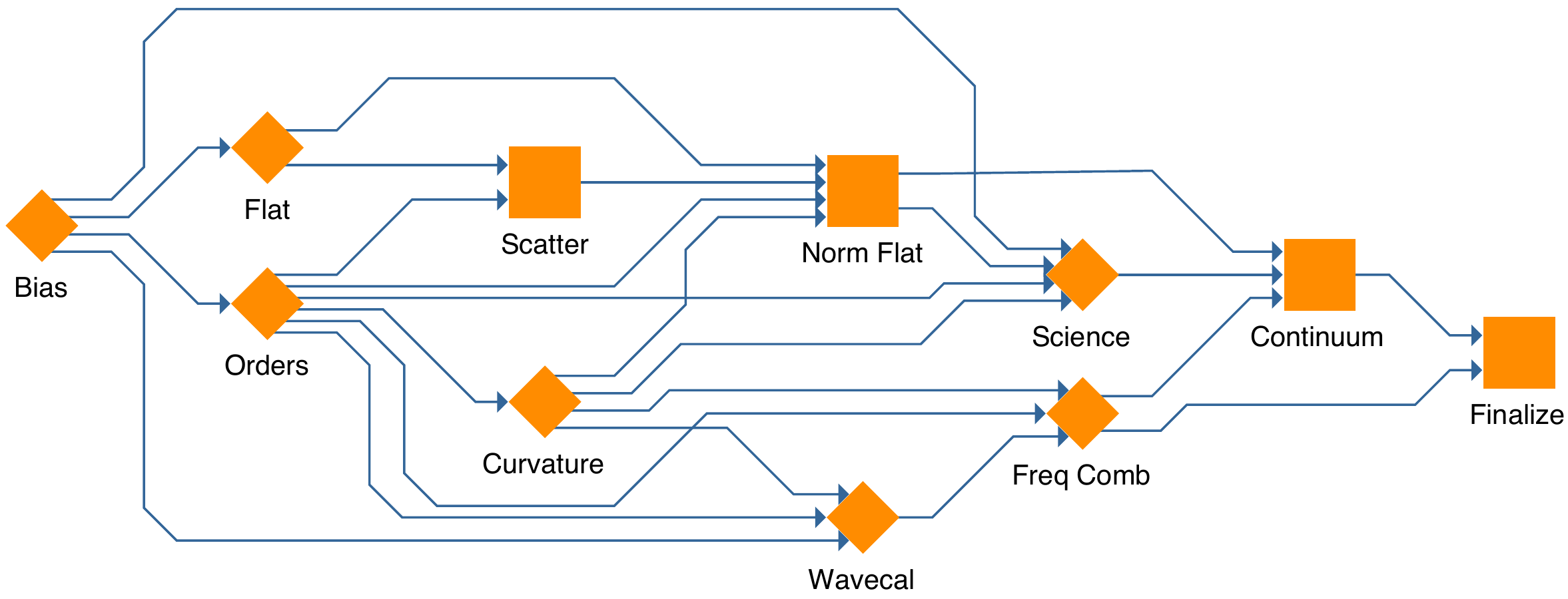}
        \caption{Connection graph of the individual components. Diamond shapes represent nodes with raw file input, while square shapes only rely on data from previous steps.}
        \label{fig:pr_graph}
    \end{figure*}

\subsubsection{What is \pyreduce?}
    \pyreduce\footnote{\url{https://github.com/AWehrhahn/PyReduce}} is a new open source implementation of the REDUCE pipeline written in Python with some C components. This new Python version is based on the existing REDUCE, which was written in IDL (Interactive Data Language). Besides the change in the language most of the code has been rewritten from scratch and new features have been added.
    Notably, a fast and speed-optimized C-version of the extraction algorithm from section \ref{sec:method} is included.
    
    The data reduction in \pyreduce is split into several individual steps, most of which produce calibration data for the science data extraction. The steps follow the methods described in the previous sections of this work, or in PAPER I, and are listed as follows (\pyreduce{} names in \textbf{bold}):
    
    \begin{enumerate}
        \item \textbf{bias} Creates the master bias frame, i.e. the intrinsic background from the detector without a light source
        \item \textbf{flat} Creates the master flat, i.e. the pixel sensitivity, from a continuum light source
        \item \textbf{orders} Traces the order locations on the detector and fits them with a polynomial
        \item \textbf{curvature} Determines the slit curvature along the orders, cf.~\autoref{sec:curvature}
        \item \textbf{scatter} Estimates the scattered light background inside the orders from the signal between orders, see section \ref{ssec:scatter}
        \item \textbf{norm\_flat} Creates the normalized flat-field from the master flat. This step also extracts the blaze functions.
        \item \textbf{wavecal} Creates the wavelength calibration, see \autoref{sec:wavecal}.
        \item \textbf{freq\_comb} Improves the wavelength solution by using a laser frequency comb (or similar). See also \autoref{sec:wavecal}.
        \item \textbf{science} Extracts the science spectrum from the science observations.
        \item \textbf{continuum} Splices together the different orders into one long spectrum, and fits the continuum level. See \autoref{sec:continuum}.
        \item \textbf{finalize} Collects all the relevant data from the different steps into the final data product and adds helpful metadata information to the FITS header.
    \end{enumerate}

\subsubsection{Using \pyreduce}
    Once \pyreduce is installed, the simplest way to use it, is by calling the \textbf{main} method. This method only requires the location of the input files, instrument, observation target, and night to start. It will find find all relevant files for this setup and perform all the steps defined above, if possible, using a predefined set of default parameters for the given instrument. These parameters are chosen to be viable in a wide range of applications, but users can of course set their own parameters.
    
    Note that to handle all kinds of different instruments, \pyreduce uses instrument specific methods, that parse the FITS headers into the \pyreduce standard format as described in \autoref{sec:method}, and identify files within the input folder. This makes it easy to apply \pyreduce to many different instruments, and even extend it for new instruments if necessary.

 The list of currently supported instruments is given in \autoref{tab:datasources}. Additional instruments can easily be added by providing the default parameters and a dictionary for the FITS header instruments-specific keywords.

         \begin{table}[ht]
        \centering
        \begin{tabular}{lrr}
             instrument & star & date \\
             UVES & HD132205 & 2010-04-01 \\
             HARPS & HD109200 & 2015-04-09 \\
             CRIRES+ &  & simulated \\
             XShooter & UX Ori & 2009-10-04 \\
             Lick APF & KIC05005618 & 2015-05-24 \\
             Keck NIRSPEC & GJ 1214 & 2010-08-05 \\
             McDonald CS23 & Vega & 2003-11-09 \\
             JWST NIRISS & GJ 436 & simulated \\
             JWST MIRI & BBR & simulated 
        \end{tabular}
        \caption{Available test datasets, which are used for investigating the results of the data reduction.}
        \label{tab:datasources}
    \end{table}

\subsubsection{Background scatter estimation}
\label{ssec:scatter}

    \begin{figure}
        \centering
        \includegraphics[width=1.\columnwidth]{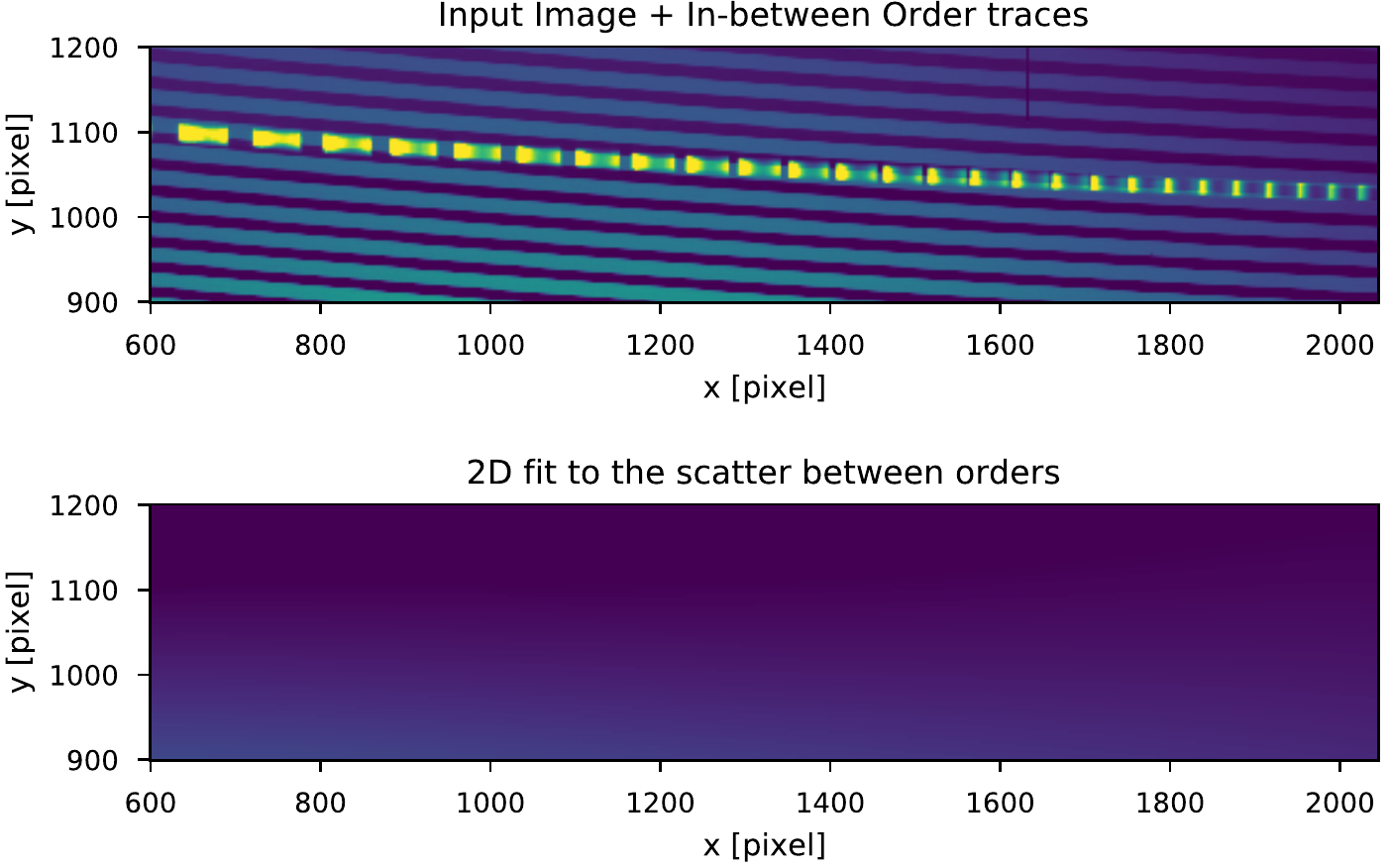}
        \caption{Scattering background in the McDonald cs23-e2 spectrograph, zoomed in on the strong ghost.
        Note how the smooth 2D model of the background is not affected by the ghost feature. Both images are on the same color scale.}
        \label{fig:scatter}
    \end{figure}

    Depending on the spectrographs properties like scattered light, it can be necessary
    to estimate this kind of background from the inter-order regions on the detector,
    and subtract it before extracting the spectra.
    The previous
    versions of REDUCE did this by extracting the order \emph{gaps} then linearly interpolating between the resulting spectra. In \pyreduce we instead perform a 2D polynomial fit to the pixel values between orders.
    This has the advantage that ghosts and other artifacts will not affect the background model.
    We illustrate this with data from the cs23-e2 spectrometer at McDonald 2.7m telescope. This instrument produces a very strong ghost image slowly crossing two spectral orders. The result of the new background estimate is shown in \autoref{fig:scatter}.

\subsection{Example application: \crires\ }
    \crires\ is an upgrade of the CRyogenic Infra-Red Echelle Spectrograph at
    ESO VLT \citep{2004SPIE.5492.1218K}. Most relevant
    in the context of this paper is the addition of a cross-disperser
    in form of a rotating wheel that carries six diffraction gratings, one for each of the YJHKLM
    bands. Thus \crires\ became a cross-dispersed echelle spectrograph
    with several spectral orders (6-10, depending on the band) registered simultaneously
    by the new, larger and better detectors (3x2048x2048, HAWAII2RG). More
    information about \crires\ can be found in \citet{2016SPIE.9908E..0ID}
    or on the ESO instrumentation web
    site\footnote{\url{https://www.eso.org/sci/facilities/develop/instruments/crires_up.html}}.
    
     Since the spectral format changed completely with the upgrade, the
    consortium and ESO re-developed the pipeline from scratch,
    keeping only a few relevant algorithms from the old CRIRES. 
    The new optical design leads to a variable tilt of the slit image
    over the focal plane, reaching in some cases as far as $\pm$4\degr\ from the vertical.
    This made obvious the need for an extraction algorithm that can handle this,
    spurring large parts of the work described in this paper.
    The slit image is close to but not exactly a straight line so we adopt a 
    parabolic model for fitting the slit image, just as described in section \ref{ssec:tilt_decomp}.

    The C-implementation of the slit-decomposition algorithm is shared between \pyreduce and the new ESO/\crires\ 
    pipeline. The routines that divide spectral orders into swaths and re-assemble the spectra are however
    written specifically for \crires, using the ESO CPL library \citep{2004SPIE.5493..444M}.

    Thereby, the slit-decomposition described in \autoref{sec:method}) is now part of the ESO framework for DRS development. Work is ongoing to include these algorithms
    into ESO's High-level Data Reduction Library (HDRL), in order to make them more easily available to other instruments.

\subsection{Example application: X-Shooter}
\begin{figure}
    \centering
    \includegraphics[width=1.\columnwidth]{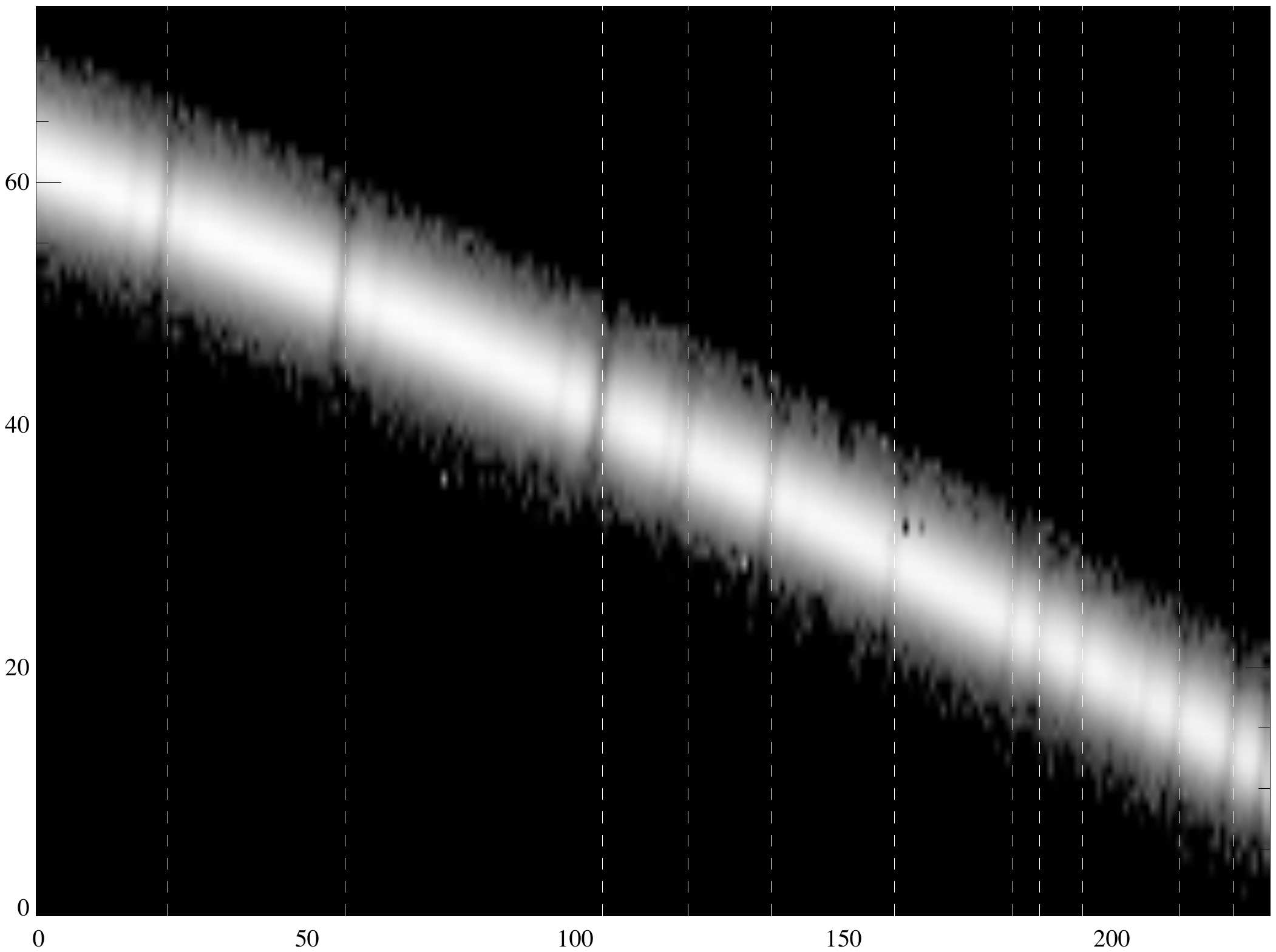}
    \caption{Selected region of an X-shooter spectrum of UX Ori (fragment of order 12 in the NIR arm).
    Vertical dashed lines mark the central position of absorption spectral features highlighting the tilt.
    The grey-scale image is shown in log scale for better visibility.}
    \label{fig:xshooter1}
\end{figure}

\begin{figure}
    \centering
    \includegraphics[width=1.\columnwidth]{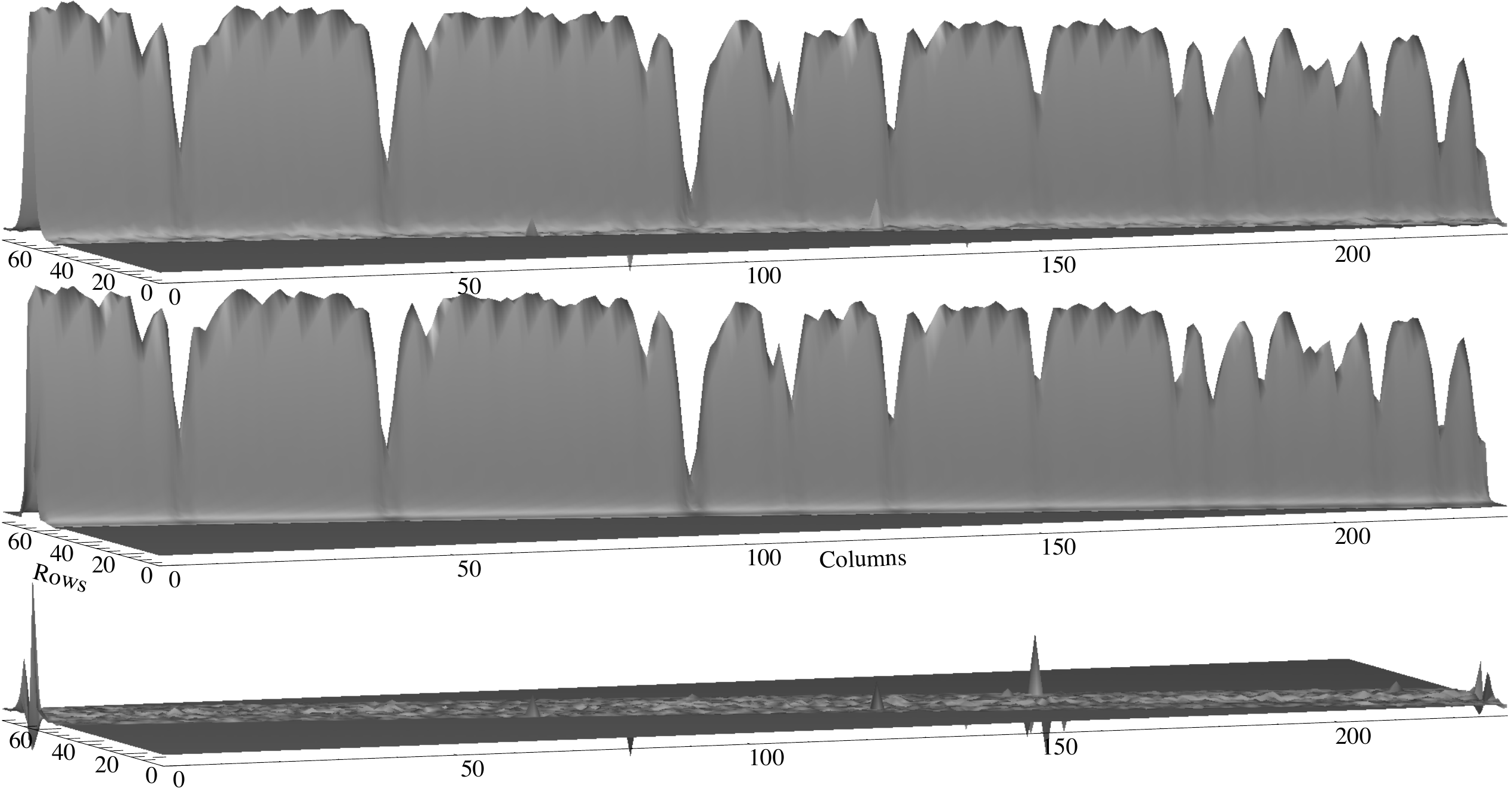}
    \caption{Comparison of the detector image $E_{x,y}$ (top) to the model $S_{x,y}$ (middle) based on tilted
    slit decomposition algorithm. The differences are shown at the bottom. Note that cosmic ray hits such
    as the one around column 150 disappear in the model.}
    \label{fig:xshooter2}
\end{figure}

\begin{figure}
    \centering
    \includegraphics[width=1.\columnwidth]{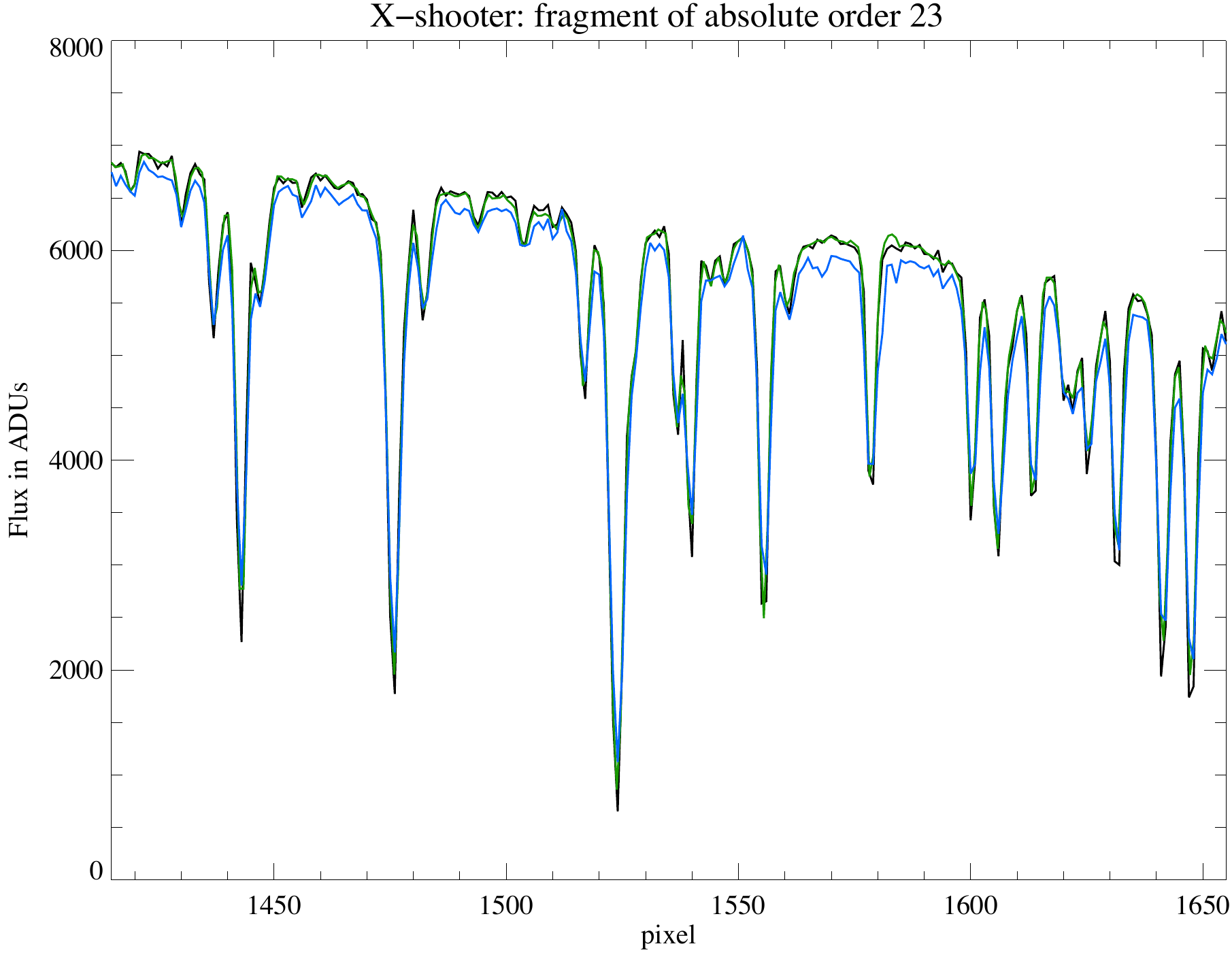}
    \caption{Comparison of extracted spectra of UX Ori obtained with three different extraction methods.
    Native X-shooter DRS optimal extraction is shown in green. Our tilted slit-decomposition extraction is
    in black and a simple vertical summation is shown in blue. Spectral lines in our extraction appear somewhat
    deeper compared to the green spectrum. The vertical summation clearly degrades spectral resolution
    and misses about 2\,\%\ of the flux.}
    \label{fig:xshooter3}
\end{figure}

    X-Shooter is a medium resolution slit spectrograph at the Very Large Telescope \citep{2011A&A...536A.105V}.
    The slit image of X-Shooter shows clear and variable curvature (see \autoref{fig:xshooter1}), which makes
    this an excellent test case for our new extraction algorithm. The native X-shooter DRS \citep{2010SPIE.7737E..28M}
    uses either interpolation for transforming spectral orders from detector coordinates into a rectangle in the
    wavelength-slit
    position plane or a 2D over-sampling of the detector pixels followed by a quasi-slit integration. Both methods
    are known to have deficiencies. Here, for the demonstration of our method we selected a random swath of 240 columns
    of order 12 in the NIR arm of a high S/N spectrum of UX Ori (ESO program ID 084.C-0952). \autoref{fig:xshooter2}
    shows a comparison of the input image and the model that is created by the extraction algorithm. Note that the
    differences between the observation and the model are small (as shown in the bottom panel) matching the noise
    estimates except for a few cosmic ray hits and a bad pixel. \autoref{fig:xshooter3} shows the extracted 1D spectrum
    for this swath in comparison with the optimal extraction result of the X-shooter DRS and a simple vertical summation.
    We see excellent agreement between the two optimal extractions with slightly higher resolution (deeper lines)
    produced by our pipeline.

\section{Conclusions}
   In the era of large and extremely large telescopes and instruments with price tags in the tens of millions of
   Euros it is important not to forget calibrations and data reduction procedures to make sure that they
   are on par with the ambitions of the coming generation of astronomical facilities. Here we describe
   previously not published algorithms and tools that address important steps in the data processing for
   modern spectroscopic instruments. 

   We have developed, implemented, and tested novel algorithms for reducing astronomical observations
   with cross-dispersed slit echelle spectrometers. The central place in this suite of algorithms is
   occupied by a slit decomposition algorithm that is capable of handling tilted and curved slit images. We have
   presented the mathematical formulation of the problem and an efficiently optimized implementation
   that is crucial due to the computationally-intensive nature of the problem. Tests and practical applications
   show excellent results in terms of preservation of the spectral resolution and the S/N of the extracted spectra.
   The algorithm is also robust against cosmic ray hits and isolated detector defects. An implementation in
   C was integrated in the ESO \crires\ DRS, ESO CPL library, and in our IDL and Python
   versions of the REDUCE package. These packages are publicly available to any interested people or
   institutions.
   
   Some of the presented algorithms can by developed further, fore example to model the PSF-asymmetry in fiber-fed
   spectrometers such as ESO HARPS and ESPRESSO  \ldots but this we leave for the next paper.

\begin{acknowledgements}
This work would not be possible without interaction within the \crires\ consortium, and with the ESO DRS group.
We are very thankful to the ESO archive that provided many of the raw and reduced data used for testing and
bench-marking for this paper. Finally, crucial financial support was provided by the Knut and Alice Wallenberg
Foundation in form of the scholarship for one of the authors and the Russian Federation grant for the project
"Study of Stars with Exoplanets" (grant 075-15-2019-1875).

\pyreduce makes use of Astropy,\footnote{http://www.astropy.org} a community-developed core Python package for Astronomy \citep{astropy:2013, astropy:2018}. 
\end{acknowledgements}

%-------------------------------------------------------------------
% References
\bibliography{main.bib}

\begin{thebibliography}{26}
\expandafter\ifx\csname natexlab\endcsname\relax\def\natexlab#1{#1}\fi

\bibitem[{{Akaike}(1974)}]{1974ITAC...19..716A}
{Akaike}, H. 1974, IEEE Transactions on Automatic Control, 19, 716

\bibitem[{{Astropy Collaboration} {et~al.}(2013){Astropy Collaboration},
  {Robitaille}, {Tollerud}, {Greenfield}, {Droettboom}, {Bray}, {Aldcroft},
  {Davis}, {Ginsburg}, {Price-Whelan}, {Kerzendorf}, {Conley}, {Crighton},
  {Barbary}, {Muna}, {Ferguson}, {Grollier}, {Parikh}, {Nair}, {Unther},
  {Deil}, {Woillez}, {Conseil}, {Kramer}, {Turner}, {Singer}, {Fox}, {Weaver},
  {Zabalza}, {Edwards}, {Azalee Bostroem}, {Burke}, {Casey}, {Crawford},
  {Dencheva}, {Ely}, {Jenness}, {Labrie}, {Lim}, {Pierfederici}, {Pontzen},
  {Ptak}, {Refsdal}, {Servillat}, \& {treicher}}]{astropy:2013}
{Astropy Collaboration}, {Robitaille}, T.~P., {Tollerud}, E.~J., {et~al.} 2013,
  \aap, 558, A33

\bibitem[{{Bolton} \& {Schlegel}(2010)}]{2010PASP..122..248B}
{Bolton}, A.~S. \& {Schlegel}, D.~J. 2010, \pasp, 122, 248

\bibitem[{{Coffinet} {et~al.}(2019){Coffinet}, {Lovis}, {Dumusque}, \&
  {Pepe}}]{2019A&A...629A..27C}
{Coffinet}, A., {Lovis}, C., {Dumusque}, X., \& {Pepe}, F. 2019, \aap, 629, A27

\bibitem[{{Cornachione} {et~al.}(2019){Cornachione}, {Bolton}, {Eastman},
  {Wilson}, {Wang}, {Johnson}, {Sliski}, {McCrady}, {Wright}, {Plavchan},
  {Johnson}, {Horner}, \& {Wittenmyer}}]{2019PASP..131l4503C}
{Cornachione}, M.~A., {Bolton}, A.~S., {Eastman}, J.~D., {et~al.} 2019, \pasp,
  131, 124503

\bibitem[{{Cui} {et~al.}(2008){Cui}, {Ye}, \& {Bai}}]{2008AcASn..49..327C}
{Cui}, B., {Ye}, Z.~F., \& {Bai}, Z.~R. 2008, Acta Astronomica Sinica, 49, 327

\bibitem[{{Cushing} {et~al.}(2004){Cushing}, {Vacca}, \&
  {Rayner}}]{2004PASP..116..362C}
{Cushing}, M.~C., {Vacca}, W.~D., \& {Rayner}, J.~T. 2004, \pasp, 116, 362

\bibitem[{{Dorn} {et~al.}(2016){Dorn}, {Follert}, {Bristow}, {Cumani},
  {Eschbaumer}, {Grunhut}, {Haimerl}, {Hatzes}, {Heiter}, {Hinterschuster},
  {Ives}, {Jung}, {Kerber}, {Klein}, {Lavail}, {Lizon}, {L{\"o}winger},
  {Molina-Conde}, {Nicholson}, {Marquart}, {Oliva}, {Origlia}, {Pasquini},
  {Paufique}, {Piskunov}, {Reiners}, {Seemann}, {Stegmeier}, {Stempels}, \&
  {Tordo}}]{2016SPIE.9908E..0ID}
{Dorn}, R.~J., {Follert}, R., {Bristow}, P., {et~al.} 2016, in Society of
  Photo-Optical Instrumentation Engineers (SPIE) Conference Series, Vol. 9908,
  \procspie, 99080I

\bibitem[{{Horne}(1986)}]{1986PASP...98..609H}
{Horne}, K. 1986, \pasp, 98, 609

\bibitem[{{Kaeufl} {et~al.}(2004){Kaeufl}, {Ballester}, {Biereichel},
  {Delabre}, {Donaldson}, {Dorn}, {Fedrigo}, {Finger}, {Fischer}, {Franza},
  {Gojak}, {Huster}, {Jung}, {Lizon}, {Mehrgan}, {Meyer}, {Moorwood}, {Pirard},
  {Paufique}, {Pozna}, {Siebenmorgen}, {Silber}, {Stegmeier}, \&
  {Wegerer}}]{2004SPIE.5492.1218K}
{Kaeufl}, H.-U., {Ballester}, P., {Biereichel}, P., {et~al.} 2004, in Society
  of Photo-Optical Instrumentation Engineers (SPIE) Conference Series, Vol.
  5492, \procspie, ed. A.~F.~M. {Moorwood} \& M.~{Iye}, 1218--1227

\bibitem[{{Kerschbaum} \& {M{\"u}ller}(2009)}]{2009AN....330..574K}
{Kerschbaum}, F. \& {M{\"u}ller}, I. 2009, Astronomische Nachrichten, 330, 574

\bibitem[{{Lenzen} {et~al.}(2005){Lenzen}, {Scherzer}, \&
  {Schindler}}]{2005A&A...443.1087L}
{Lenzen}, F., {Scherzer}, O., \& {Schindler}, S. 2005, \aap, 443, 1087

\bibitem[{{Marsh}(1989)}]{1989PASP..101.1032M}
{Marsh}, T.~R. 1989, \pasp, 101, 1032

\bibitem[{{Mayor} {et~al.}(2003){Mayor}, {Pepe}, {Queloz}, {Bouchy},
  {Rupprecht}, {Lo Curto}, {Avila}, {Benz}, {Bertaux}, {Bonfils}, {Dall},
  {Dekker}, {Delabre}, {Eckert}, {Fleury}, {Gilliotte}, {Gojak}, {Guzman},
  {Kohler}, {Lizon}, {Longinotti}, {Lovis}, {Megevand}, {Pasquini}, {Reyes},
  {Sivan}, {Sosnowska}, {Soto}, {Udry}, {van Kesteren}, {Weber}, \&
  {Weilenmann}}]{2003Msngr.114...20M}
{Mayor}, M., {Pepe}, F., {Queloz}, D., {et~al.} 2003, The Messenger, 114, 20

\bibitem[{{McKay} {et~al.}(2004){McKay}, {Ballester}, {Banse}, {Izzo}, {Jung},
  {Kiesgen}, {Kornweibel}, {Lundin}, {Modigliani}, {Palsa}, \&
  {Sabet}}]{2004SPIE.5493..444M}
{McKay}, D.~J., {Ballester}, P., {Banse}, K., {et~al.} 2004, in Society of
  Photo-Optical Instrumentation Engineers (SPIE) Conference Series, Vol. 5493,
  \procspie, ed. P.~J. {Quinn} \& A.~{Bridger}, 444--452

\bibitem[{{Milakovi{\'c}} {et~al.}(2020){Milakovi{\'c}}, {Pasquini}, {Webb}, \&
  {Lo Curto}}]{2020MNRAS.493.3997M}
{Milakovi{\'c}}, D., {Pasquini}, L., {Webb}, J.~K., \& {Lo Curto}, G. 2020,
  \mnras, 493, 3997

\bibitem[{{Modigliani} {et~al.}(2010){Modigliani}, {Goldoni}, {Royer},
  {Haigron}, {Guglielmi}, {Fran{\c{c}}ois}, {Horrobin}, {Bristow}, {Vernet},
  {Moehler}, {Kerber}, {Ballester}, {Mason}, \&
  {Christensen}}]{2010SPIE.7737E..28M}
{Modigliani}, A., {Goldoni}, P., {Royer}, F., {et~al.} 2010, in Society of
  Photo-Optical Instrumentation Engineers (SPIE) Conference Series, Vol. 7737,
  \procspie, 773728

\bibitem[{{Petersburg} {et~al.}(2020){Petersburg}, {Ong}, {Zhao}, {Blackman},
  {Brewer}, {Buchhave}, {Cabot}, {Davis}, {Jurgenson}, {Leet}, {McCracken},
  {Sawyer}, {Sharov}, {Tronsgaard}, {Szymkowiak}, \&
  {Fischer}}]{2020AJ....159..187P}
{Petersburg}, R.~R., {Ong}, J.~M.~J., {Zhao}, L.~L., {et~al.} 2020, \aj, 159,
  187

\bibitem[{{Piskunov} \& {Valenti}(2002)}]{2002A&A...385.1095P}
{Piskunov}, N.~E. \& {Valenti}, J.~A. 2002, \aap, 385, 1095

\bibitem[{{Price-Whelan} {et~al.}(2018){Price-Whelan}, {Sip{\H{o}}cz},
  {G{\"u}nther}, {Lim}, {Crawford}, {Conseil}, {Shupe}, {Craig}, {Dencheva},
  {Ginsburg}, {VanderPlas}, {Bradley}, {P{\'e}rez-Su{\'a}rez}, {de Val-Borro},
  {Paper Contributors}, {Aldcroft}, {Cruz}, {Robitaille}, {Tollerud},
  {Coordination Committee}, {Ardelean}, {Babej}, {Bach}, {Bachetti}, {Bakanov},
  {Bamford}, {Barentsen}, {Barmby}, {Baumbach}, {Berry}, {Biscani}, {Boquien},
  {Bostroem}, {Bouma}, {Brammer}, {Bray}, {Breytenbach}, {Buddelmeijer},
  {Burke}, {Calderone}, {Cano Rodr{\'\i}guez}, {Cara}, {Cardoso}, {Cheedella},
  {Copin}, {Corrales}, {Crichton}, {D{\textquoteright}Avella}, {Deil},
  {Depagne}, {Dietrich}, {Donath}, {Droettboom}, {Earl}, {Erben}, {Fabbro},
  {Ferreira}, {Finethy}, {Fox}, {Garrison}, {Gibbons}, {Goldstein}, {Gommers},
  {Greco}, {Greenfield}, {Groener}, {Grollier}, {Hagen}, {Hirst}, {Homeier},
  {Horton}, {Hosseinzadeh}, {Hu}, {Hunkeler}, {Ivezi{\'c}}, {Jain}, {Jenness},
  {Kanarek}, {Kendrew}, {Kern}, {Kerzendorf}, {Khvalko}, {King}, {Kirkby},
  {Kulkarni}, {Kumar}, {Lee}, {Lenz}, {Littlefair}, {Ma}, {Macleod},
  {Mastropietro}, {McCully}, {Montagnac}, {Morris}, {Mueller}, {Mumford},
  {Muna}, {Murphy}, {Nelson}, {Nguyen}, {Ninan}, {N{\"o}the}, {Ogaz}, {Oh},
  {Parejko}, {Parley}, {Pascual}, {Patil}, {Patil}, {Plunkett}, {Prochaska},
  {Rastogi}, {Reddy Janga}, {Sabater}, {Sakurikar}, {Seifert}, {Sherbert},
  {Sherwood-Taylor}, {Shih}, {Sick}, {Silbiger}, {Singanamalla}, {Singer},
  {Sladen}, {Sooley}, {Sornarajah}, {Streicher}, {Teuben}, {Thomas},
  {Tremblay}, {Turner}, {Terr{\'o}n}, {van Kerkwijk}, {de la Vega}, {Watkins},
  {Weaver}, {Whitmore}, {Woillez}, {Zabalza}, \& {Contributors}}]{astropy:2018}
{Price-Whelan}, A.~M., {Sip{\H{o}}cz}, B.~M., {G{\"u}nther}, H.~M., {et~al.}
  2018, \aj, 156, 123

\bibitem[{{Quirrenbach} {et~al.}(2016){Quirrenbach}, {Amado}, {Caballero},
  {Mundt}, {Reiners}, {Ribas}, {Seifert}, {Abril}, {Aceituno},
  {Alonso-Floriano}, {Anwand -Heerwart}, {Azzaro}, {Bauer}, {Barrado},
  {Becerril}, {Bejar}, {Benitez}, {Berdinas}, {Brinkm{\"o}ller}, {Cardenas},
  {Casal}, {Claret}, {Colom{\'e}}, {Cortes-Contreras}, {Czesla}, {Doellinger},
  {Dreizler}, {Feiz}, {Fernandez}, {Ferro}, {Fuhrmeister}, {Galadi},
  {Gallardo}, {G{\'a}lvez-Ortiz}, {Garcia-Piquer}, {Garrido}, {Gesa},
  {G{\'o}mez Galera}, {Gonz{\'a}lez Hern{\'a}ndez}, {Gonzalez Peinado},
  {Gr{\"o}zinger}, {Gu{\`a}rdia}, {Guenther}, {de Guindos}, {Hagen}, {Hatzes},
  {Hauschildt}, {Helmling}, {Henning}, {Hermann}, {Hern{\'a}ndez Arabi},
  {Hern{\'a}ndez Casta{\~n}o}, {Hern{\'a}ndez Hernando}, {Herrero}, {Huber},
  {Huber}, {Huke}, {Jeffers}, {de Juan}, {Kaminski}, {Kehr}, {Kim}, {Klein},
  {Kl{\"u}ter}, {K{\"u}rster}, {Lafarga}, {Lara}, {Lamert}, {Laun},
  {Launhardt}, {Lemke}, {Lenzen}, {Llamas}, {Lopez del Fresno},
  {L{\'o}pez-Puertas}, {L{\'o}pez-Santiago}, {Lopez Salas}, {Magan
  Madinabeitia}, {Mall}, {Mandel}, {Mancini}, {Marin Molina}, {Maroto
  Fern{\'a}ndez}, {Mart{\'\i}n}, {Mart{\'\i}n-Ruiz}, {Marvin}, {Mathar},
  {Mirabet}, {Montes}, {Morales}, {Morales Mu{\~n}oz}, {Nagel}, {Naranjo},
  {Nowak}, {Palle}, {Panduro}, {Passegger}, {Pavlov}, {Pedraz}, {Perez},
  {P{\'e}rez-Medialdea}, {Perger}, {Pluto}, {Ram{\'o}n}, {Rebolo}, {Redondo},
  {Reffert}, {Reinhart}, {Rhode}, {Rix}, {Rodler}, {Rodr{\'\i}guez},
  {Rodr{\'\i}guez L{\'o}pez}, {Rohloff}, {Rosich}, {Sanchez Carrasco},
  {Sanz-Forcada}, {Sarkis}, {Sarmiento}, {Sch{\"a}fer}, {Schiller}, {Schmidt},
  {Schmitt}, {Sch{\"o}fer}, {Schweitzer}, {Shulyak}, {Solano}, {Stahl},
  {Storz}, {Tabernero}, {Tala}, {Tal-Or}, {Ulbrich}, {Veredas}, {Vico Linares},
  {Vilardell}, {Wagner}, {Winkler}, {Zapatero Osorio}, {Zechmeister},
  {Ammler-von Eiff}, {Anglada-Escud{\'e}}, {del Burgo}, {Garcia-Vargas},
  {Klutsch}, {Lizon}, {Lopez-Morales}, {Ofir}, {P{\'e}rez-Calpena}, {Perryman},
  {S{\'a}nchez-Blanco}, {Strachan}, {St{\"u}rmer}, {Su{\'a}rez}, {Trifonov},
  {Tulloch}, \& {Xu}}]{2016SPIE.9908E..12Q}
{Quirrenbach}, A., {Amado}, P.~J., {Caballero}, J.~A., {et~al.} 2016, in
  Society of Photo-Optical Instrumentation Engineers (SPIE) Conference Series,
  Vol. 9908, Ground-based and Airborne Instrumentation for Astronomy VI, 990812

\bibitem[{Tikhonov \& Arsenin(1977)}]{Tikhonov1977}
Tikhonov, A.~N. \& Arsenin, V.~Y. 1977, Solutions of ill-posed problems
  (Washington, D.C.: John Wiley \& Sons, New York: V. H. Winston \& Sons),
  xiii+258, translated from the Russian, Preface by translation editor Fritz
  John, Scripta Series in Mathematics

\bibitem[{{Valenti}(1994)}]{1994PhDT........16V}
{Valenti}, J.~A. 1994, PhD thesis, UNIVERSITY OF CALIFORNIA, BERKELEY.

\bibitem[{{Vernet} {et~al.}(2011){Vernet}, {Dekker}, {D'Odorico}, {Kaper},
  {Kjaergaard}, {Hammer}, {Randich}, {Zerbi}, {Groot}, {Hjorth}, {Guinouard},
  {Navarro}, {Adolfse}, {Albers}, {Amans}, {Andersen}, {Andersen}, {Binetruy},
  {Bristow}, {Castillo}, {Chemla}, {Christensen}, {Conconi}, {Conzelmann},
  {Dam}, {de Caprio}, {de Ugarte Postigo}, {Delabre}, {di Marcantonio},
  {Downing}, {Elswijk}, {Finger}, {Fischer}, {Flores}, {Fran{\c{c}}ois},
  {Goldoni}, {Guglielmi}, {Haigron}, {Hanenburg}, {Hendriks}, {Horrobin},
  {Horville}, {Jessen}, {Kerber}, {Kern}, {Kiekebusch}, {Kleszcz}, {Klougart},
  {Kragt}, {Larsen}, {Lizon}, {Lucuix}, {Mainieri}, {Manuputy}, {Martayan},
  {Mason}, {Mazzoleni}, {Michaelsen}, {Modigliani}, {Moehler}, {M{\o}ller},
  {Norup S{\o}rensen}, {N{\o}rregaard}, {P{\'e}roux}, {Patat}, {Pena}, {Pragt},
  {Reinero}, {Rigal}, {Riva}, {Roelfsema}, {Royer}, {Sacco}, {Santin},
  {Schoenmaker}, {Spano}, {Sweers}, {Ter Horst}, {Tintori}, {Tromp}, {van
  Dael}, {van der Vliet}, {Venema}, {Vidali}, {Vinther}, {Vola}, {Winters},
  {Wistisen}, {Wulterkens}, \& {Zacchei}}]{2011A&A...536A.105V}
{Vernet}, J., {Dekker}, H., {D'Odorico}, S., {et~al.} 2011, \aap, 536, A105

\bibitem[{{von Littrow}(1863)}]{Littrow}
{von Littrow}, O. 1863, Sitzungsberichte der k. Akad. d. Wiss.,
  mathem.-naturwiss. Klasse, XLVII, Band. 47 Abt. 2

\bibitem[{{Zechmeister} {et~al.}(2014){Zechmeister}, {Anglada-Escud{\'e}}, \&
  {Reiners}}]{2014A&A...561A..59Z}
{Zechmeister}, M., {Anglada-Escud{\'e}}, G., \& {Reiners}, A. 2014, \aap, 561,
  A59

\end{thebibliography}
\end{document}